\documentclass[aps,prd,showpacs,twocolumn,floatfix,nofootinbib]{revtex4-1}
\usepackage[dvipdfmx]{graphicx}
\usepackage{amsmath,amssymb,siunitx}
\usepackage{color,ulem}
\usepackage{graphicx}
\usepackage{bm,latexsym,amsmath,amssymb,amsfonts,mathrsfs}
\usepackage{dcolumn}   
\usepackage{color}

\newcommand{\simgt}{\lower.5ex\hbox{$\; \buildrel > \over \sim \;$}}
\newcommand{\simlt}{\lower.5ex\hbox{$\; \buildrel < \over \sim \;$}}
\newcommand{\solM}{M_{\odot}}

\begin{document}
\title{Follow-up analyses of the binary-neutron-star signals GW170817 and GW190425 by using post-Newtonian waveform models}
\author{Tatsuya Narikawa$^{1}$}
\email{narikawa@icrr.u-tokyo.ac.jp}
\author{Nami Uchikata$^{1}$}
\email{uchikata@icrr.u-tokyo.ac.jp}
\affiliation{
$^1$Institute for Cosmic Ray Research, The University of Tokyo, Chiba 277-8582, Japan\\
}
\date{\today}

\begin{abstract}
We reanalyze the binary-neutron-star signals, GW170817 and GW190425, 
focusing on the inspiral regime. 
We use post-Newtonian waveform models  
as templates, which are theoretically rigid and efficiently describe the inspiral regime.
We study potential systematic difference in estimates of the binary tidal deformability $\tilde{\Lambda}$ 
by using different descriptions for the point-particle dynamics and tidal effects.
We find that the estimates of $\tilde{\Lambda}$ show no significant systematic difference among 
three models for the point-particle parts: \texttt{TF2}, \texttt{TF2g}, and \texttt{TF2+},
when they employ the same tidal model.
We compare different tidal descriptions given by different post-Newtonian orders in the tidal phase.
Our results indicate that the estimates of $\tilde{\Lambda}$ slightly depend on the post-Newtonian order in the tidal phase and 
an increase in the tidal post-Newtonian order does not lead to a monotonic change in the estimate of $\tilde{\Lambda}$.
We also compare the estimate of $\tilde{\Lambda}$ obtained by the post-Newtonian tidal model and numerical-relativity calibrated tidal models.
We find that the post-Newtonian model gives slightly larger estimate of $\tilde{\Lambda}$ and wider posterior distribution than the numerical-relativity calibrated models.
According to Bayesian model comparison, it is difficult to identify a preference among the post-Newtonian orders by relying on the GW170817 and GW190425 data.
Our results indicate no preference among numerical-relativity calibrated tidal models over the post-Newtonian model.
Additionally, we present constraints on equation-of-state models for neutron stars with the post-Newtonian model, 
which show that the GW170817 data disfavor less compact models,
though they are slightly weaker constraints than the numerical-relativity calibrated tidal models.
\end{abstract}


\maketitle

\section{Introduction}
\label{sec:Introduction}
The post-Newtonian (PN) formalism is theoretically rigid 
and can efficiently describe the inspiral regime of gravitational waves (GWs) for compact binary coalescences~\cite{Blanchet:2013haa, PoissonWillGraivty, Isoyama:2020lls}.
The formalism is an approximation of general relativity, which is valid under the assumption of slow motion and weak field.
GWs provide the unique laboratory of extremely high density objects 
through extraction of tidal effects~\cite{Lattimer:2015nhk, Baiotti:2019sew, Dietrich:2020eud, Chatziioannou:2020pqz, Ozel:2016oaf}.
The tidal deformability of neutron star (NS) matter can quantify tidal effects and
characterize equation-of-state (EOS) models for NSs~\cite{Hinderer:2007mb, Damour:2009vw, Postnikov:2010yn}.
For binary neutron stars (BNSs), 
closer to the merger, corrections due to the finite-size effects of NSs become important,
which can be described by the tidal deformability~\cite{Flanagan:2007ix, Hinderer:2009ca, Vines:2011ud}.
PN expansion gives us conservative constraints on the tidal deformability and EOS models for NSs.
PN tidal waveform (hereafter we call it \texttt{PNTidal}) have been derived up to 2.5PN (relative 5+2.5PN) order for phase~\cite{Damour:2012yf, Bini:2012gu, Agathos:2015uaa} and have used in analyses of the BNS signals detected by the Advanced LIGO and Advanced Virgo detectors, GW170817 and/or GW190425~\cite{TheLIGOScientific:2017qsa, Abbott:2018wiz, De:2018uhw, LIGOScientific:2018mvr, Abbott:2020uma, LIGOScientific:2020ibl, Dai:2018dca, Narikawa:2018yzt, Narikawa:2019xng, Gamba:2020wgg, Breschi:2021wzr} (see also Refs.~\cite{LIGOScientific:2018ehx, Pratten:2019sed, Pan:2020tht}).
Some results have shown the tendency that the \texttt{PNTidal} model gives weaker constraint on the binary tidal deformability, $\tilde{\Lambda}$, than the numerical-relativity (NR) calibrated models (e.g.,~\cite{Abbott:2018wiz, Narikawa:2019xng, Gamba:2020wgg}).

Recently, the complete and correct PN tidal phase up to 5+2.5PN order
for mass quadrupole, mass octupole, and current quadrupole interactions have been derived
by the PN-matched multiplolar-post-Minkowskian formalism~\cite{Henry:2020ski}.
In our previous study~\cite{Narikawa:2021pak}, 
we rewrite the the complete and correct form 
for the mass quadrupole interactions as a function of the dimensionless tidal deformability for the individual stars, 
$\Lambda$, in a convenient way for data analyses.
And we have reanalyzed the data around low-mass events identified as binary black holes (BBHs) by using the corrected version of the \texttt{PNTidal} model
to test the exotic compact object hypothesis.
However, the corrected \texttt{PNTidal} model has not been used in analyses of BNS signals yet, except for analyses in Ref.~\cite{Breschi:2021wzr}.
In Ref.~\cite{Breschi:2021wzr}, they applied the \texttt{BAJES} pipeline, which has implemented 
the correct version of the \texttt{PNTidal}, to GW170817.

Many signal injection studies have investigated waveform systematic biases 
in the inference of $\tilde{\Lambda}$ from injected BNS signals.
They have found that 
for the BNS signals with signal-to-noise ratio (SNR) higher than 80,
waveform systematics may introduce biases in the inference of the injected $\tilde{\Lambda}$
~\cite{Dudi:2018jzn, Samajdar:2018dcx, Messina:2019uby, Samajdar:2019ulq, Agathos:2019sah, Gamba:2020wgg}
and in the inference of EOS models for NSs~\cite{Gamba:2020wgg, Pratten:2021pro, Chatziioannou:2021tdi, Kunert:2021hgm}.
Some of them have implied when the \texttt{TEOBResumS} waveform~\cite{Nagar:2018zoe, Nagar:2018plt, Akcay:2018yyh} or hybridized waveforms with NR waveforms are injected,
the "\texttt{IMRPhenom}" BBH baseline employing the \texttt{NRTidal}~\cite{Dietrich:2017aum} model as a tidal part
systematically underestimates $\tilde{\Lambda}$,
while the \texttt{TF2\_PNTidal} model yields overall consistent estimates.
Reference~\cite{Gamba:2020wgg} has discussed that this is understandable from that the difference in the point-mass and tidal sector between the models have opposite sign and partially compensate each other.
Here, 
\texttt{TF2} is the abbreviation of \texttt{TaylorF2}, 
which is the PN waveform model for a point-particle part~\cite{Dhurandhar:1992mw, Buonanno:2009zt, Blanchet:2013haa}.
The above "\texttt{IMRPhenom}" series are phenomenological constructed inspiral-merger-ringdown models for the BBH baseline part for convenience in data analyses, 
represented by \texttt{IMRPhenomD}, which is a model for spins (anti) aligned to the orbital angular momentum ~\cite{Khan:2015jqa, Husa:2015iqa}, 
and \texttt{IMRPhenomPv2}, which is a spin precessing model~\cite{Hannam:2013oca, Schmidt:2012rh, Schmidt:2014iyl}.
The \texttt{TEOBResumS} model is based on the effective-one-body (EOB) description of the general relativistic two-body problem~\cite{Buonanno:1998gg, Buonanno:2000ef} for (anti) aligned spin compact binary coalescences. The model is extended to the BNS coalescences. 
The \texttt{NRTidal} model is a NR simulation calibrated tidal model.

Waveform systematics in the inference of tidal parameters from BNS signals GW170817 and/or GW190425
have been investigated for the point-particle part as well as tidal effects
~\cite{Abbott:2018wiz, Narikawa:2019xng, Gamba:2020wgg, Ashton:2021yum}.
Basically, all studies have shown the agreement in the estimate of $\tilde{\Lambda}$ for GW170817 and GW190425 between the different waveform models employed in the inference under a similar setup.
This is mainly because for GW170817, with its SNR of $\sim30$, waveform systematics are within the statistical error.
However, a closer look at the results reveals that the \texttt{NRTidal} and \texttt{NRTidalv2}~\cite{Dietrich:2019kaq} models give slightly smaller $\tilde{\Lambda}$ and tighter constraints for GW170817 than the \texttt{TF2\_PNTidal} and \texttt{TEOBResumS} models.
While this is simply a consequence of the phase shift, taking into account the results of the BNS signal injection studies, the NR calibrated models might bias on the estimates of $\tilde{\Lambda}$ for GW170817.
References~\cite{Dai:2018dca, Narikawa:2019xng, Gamba:2020wgg} have shown that the inspiral-only analyses with the upper frequency cutoff $f_\mathrm{high}=1000~\mathrm{Hz}$ give larger $\tilde{\Lambda}$ and weaker constraints than the full analyses with $f_\mathrm{high}=2048~\mathrm{Hz}$. 
Reference~\cite{Narikawa:2019xng} has shown that for the \texttt{NRTidalv2} model, $\tilde{\Lambda}=445_{-330}^{+412}$ for $f_\mathrm{high}=1000~\mathrm{Hz}$ and $312_{-208}^{+498}$ for $f_\mathrm{high}=2048~\mathrm{Hz}$. 
The \texttt{KyotoTidal} model~\cite{Kawaguchi:2018gvj} also gives smaller estimate of $\tilde{\Lambda}$ and tighter constraint as for the \texttt{NRTidalv2} model than the \texttt{PNTidal} model~\cite{Narikawa:2019xng}.
Reference~\cite{Gamba:2020wgg} gives the following result: the marginalized posterior distribution of $\tilde{\Lambda}$ measured with the \texttt{TEOBResumS} model is comparable with the \texttt{TF2\_PNTidal} model for $f_\mathrm{high}=1000~\mathrm{Hz}$ is also interesting.
Here, the \texttt{NRTidalv2} model is an upgrade of the \texttt{NRTidal} model
and the \texttt{KyotoTidal} model is another NR calibrated model for the inspiral regime of BNS mergers.

Model comparison among waveform models has been investigated on GW170817 and GW190425.
Reference~\cite{Gamba:2020wgg} has shown that the values of Bayesian evidences
indicate that it is not possible to identify a preferred waveform model exclusively on the GW170817 data
by comparing different waveform models: \texttt{TEOBResumS}, \texttt{TF2\_PNTidal}, and \texttt{NRTidal}.
Reference~\cite{Ashton:2021yum} has applied a hypermodel approach to analyze GW170817, GW190425, and the subthreshold candidate GW200311\_103121.
The method finds that the \texttt{TEOBResumS} model is the most successful at predicting the GW170817 data.
Although the odds do not exceed the threshold of a significant preference to the \texttt{TEOBResumS},
they stress that the mild preference to the \texttt{TEOBResumS} model over the \texttt{NRTidalv2} and \texttt{SEOBNRv4T}~\cite{Hinderer:2016eia, Steinhoff:2016rfi} models is worthy of further investigation.

Some studies have investigated constraints on EOS models for NSs by using information obtained from GW170817~\cite{TheLIGOScientific:2017qsa, Abbott:2018wiz, De:2018uhw, Abbott:2018exr, LIGOScientific:2019eut, Landry:2018prl, Landry:2020vaw, Capano:2019eae, Narikawa:2019xng, Pacilio:2021jmq}.
They have found that it is difficult to rule out the majority of theoretical models,
while less compact models are disfavored.
Furthermore, it is also difficult to rule out the possibility that GW170817 was a BBH or BH-NS coalescence by the GW data alone~\cite{LIGOScientific:2019eut}.

In this paper, we reanalyze the data around two BNS signals, GW170817 and GW190425,
focusing on the inspiral regime with the frequency up to $1000~\mathrm{Hz}$. 
We mainly use PN waveform models 
as templates here, which are theoretically rigid and efficiently describe the inspiral regime,
while we focused on NR calibrated waveform models in our previous study~\cite{Narikawa:2019xng}.
We study potential systematic difference in estimates of $\tilde{\Lambda}$ for GW170817 and GW190425 by using different descriptions for the point-particle dynamics and tidal effects.
First, we compare the estimates of $\tilde{\Lambda}$ among 
three post-Newtonian models for the point-particle parts: the \texttt{TF2}, \texttt{TF2g}, and \texttt{TF2+} models.
Here, the \texttt{TF2g} model~\cite{Messina:2019uby} and the \texttt{TF2+} model~\cite{Kawaguchi:2018gvj} are phenomenological extended models of the \texttt{TF2} model.
Then, we compare different tidal phase descriptions from different PN orders in the tidal phase up to 5PN, 5+1PN, 5+1.5PN, 5+2PN, and 5+2.5PN orders.
By analyzing GW170817 and GW190425, 
we examine a systematic difference in estimates of $\tilde{\Lambda}$ among them. 
For comparison, we discuss a systematic difference in the estimates of $\tilde{\Lambda}$ for GW170817 
between the PN and NR calibrated models and examine a preference to the PN model over the NR calibrated models by Bayesian model comparison.
Finally, we also present the constraints on EOS models for NSs with the corrected PN tidal waveform model.

The remainder of this paper is organized as follows.
In Sec.~\ref{sec:PE}, we explain the methods of Bayesian parameter estimation for GWs 
including waveform models used in our analyses.
In Sec.~\ref{sec:results}, we mainly present results of our analyses of GW170817 and GW190425
by using the \texttt{TF2+\_PNTidal} waveform model.
We compare the results
by using the \texttt{TF2}, \texttt{TF2g}, and \texttt{TF2+} waveform models for a point-particle part.
We examine the effect of the different PN orders in the tidal phase.
We discuss a systematic difference between the PN and NR calibrated models for tidal effects.
We also present the constraints on EOS models for NSs.
Section~\ref{sec:summary} is devoted to a conclusion.
In Appendix~\ref{sec:Old_vs_Corrected_PNTidal}, we present phase difference between the old and corrected versions of the \texttt{PNTidal} model.
In Appendix~\ref{sec:WFsys}, we present estimates of source parameters for completeness obtained by using three models for a point-particle part: the \texttt{TF2}, \texttt{TF2g}, and \texttt{TF2+} models.

We employ the units $c=G=1$, where $c$ and $G$ are the speed of light and the gravitational constant, respectively.

\section{Parameter estimation methods}
\label{sec:PE}

\subsection{Tidal deformability}
\label{sec:tidal}
There are several features of NSs that depend on EOSs (see Refs.~\cite{Lattimer:2015nhk, Baiotti:2019sew, Dietrich:2020eud, Chatziioannou:2020pqz} for review).
In this paper, we focus on the tidal deformability.

In inspiraling BNSs, at the leading order,
the tidally induced quadrupole moment tensor $Q_{ij,\mathrm{Tidal}}$ is proportional to 
the companion's tidal field $\mathcal{E}_{ij}$ as $Q_{ij,\mathrm{Tidal}} = -\lambda \mathcal{E}_{ij}$.
The information about the EOS (or structure) can be quantified by the tidal deformability parameter 
$\lambda$~\cite{Flanagan:2007ix}.
The leading-order tidal contribution to the GW phase evolution (relative 5PN order) arises
through the symmetric contribution of  tidal deformation, the binary tidal deformability~\cite{Flanagan:2007ix, Hinderer:2007mb, Vines:2011ud}
\begin{eqnarray}
 \tilde{\Lambda} = \frac{16}{13} \left[ 
 \left( 1+11 X_2 \right) X_1^4 \Lambda_1 + \left( 1+11 X_1 \right) X_2^4 \Lambda_2 
\right],
\end{eqnarray}
which is a mass-weighted linear combination of the both component tidal parameters,
with $X_{1,2} = m_{1,2} / M$ 
where $m_{1,2}$ is the component mass, $M=m_1+m_2$ is the total mass, and 
$\Lambda_{1,2}$ is the dimensionless mass quadrupole tidal deformability parameter of each object 
defined as $\Lambda_{1,2}=\lambda_{1,2}/m_{1,2}^5$.
The tidal effects to the GW phase are dominated by the symmetric contributions of $\tilde{\Lambda}$
and the antisymmetric contributions of $\delta \tilde{\Lambda}$,
\begin{eqnarray}
 \delta\tilde{\Lambda} =  
 \left( X_1^2 - \frac{7996}{1319} X_1 X_2 - \frac{11005}{1319} X_2^2 \right) X_1^4 \Lambda_1 - (1 \leftrightarrow 2), \nonumber \\
\end{eqnarray}
are always subdominant for realistic cases~\cite{Favata:2013rwa, Wade:2014vqa}.
For the equal NSs with the same mass ($m_1=m_2$) and identical tidal deformability ($\Lambda_1=\Lambda_2=\Lambda$),
$\tilde{\Lambda} \rightarrow \Lambda$ and $\delta\tilde{\Lambda} \rightarrow 0$.
The tidal deformability characterize the compact objects.
For NSs, $\Lambda$ is $\sim100 - 2000$ for $m\sim1.4~\solM$, depending on the EOS models for NSs~\cite{Hinderer:2009ca, Postnikov:2010yn, Vines:2011ud, Damour:2012yf, Lattimer:2015nhk, LIGOScientific:2019eut, Chatziioannou:2020pqz}.
The upper bound on the binary tidal deformability $\tilde{\Lambda}$ by GW170817 is about 800 by using the \texttt{IMRPhenomPv2\_NRTidal} model for the low-spin prior and the upper frequency cutoff $2048~\mathrm{Hz}$~\cite{Abbott:2018wiz} (see also Refs.~\cite{Narikawa:2018yzt, Narikawa:2019xng} for reanalysis). 
The constraint can be further improved~\cite{Abbott:2018exr, LIGOScientific:2019eut} by assuming the EOS to be common for both NS~\cite{Yagi:2015pkc, Yagi:2016qmr, Chatziioannou:2018vzf}
(but see also Ref.~\cite{Kastaun:2019bxo} for systematic effects due to assuming the EOS-independent relations).
$\Lambda$ for BHs in classical GR vanishes as shown for a Schwarzschild BH~\cite{Binnington:2009bb, Damour:2009wj}
and for a Kerr BH~\cite{Poisson:2014gka, Pani:2015hfa, Landry:2015zfa, LeTiec:2020spy, Chia:2020yla, Goldberger:2020fot, Charalambous:2021mea}.

\subsection{Waveform models for inspiraling binary neutron stars}
\label{sec:WF}

We use the PN waveform models, summarized in Table \ref{table:WFs}, 
which can efficiently describe the inspiral regime of GWs for compact binary coalescences~\cite{Blanchet:2013haa, PoissonWillGraivty, Isoyama:2020lls}.
The frequency-domain gravitational waveform for BNSs can be written as
\begin{eqnarray}
 \tilde{h}_\mathrm{BNS}(f) = \left( A_\mathrm{BBH}(f) + A_\mathrm{Tidal}(f) \right) e^{i \left( \Psi_\mathrm{BBH}(f) + \Psi_\mathrm{Tidal}(f) \right)}, \nonumber \\
\end{eqnarray}
where both amplitude and phase of the GW signal consist of the BBH part: $A_\mathrm{BBH}(f)$ and $\Psi_\mathrm{BBH}(f)$, and the additional tidal effects: $A_\mathrm{Tidal}(f)$ and $\Psi_\mathrm{Tidal}(f)$.
The PN amplitude formula for $A_\mathrm{BBH} (f)$ is up to the 3PN order as summarized in Ref.~\cite{Khan:2015jqa}, 
where the point particle and the spin effects are included.
The leading-order term is written as approximately
$\sim D_\mathrm{L}^{-1} (\mathcal{M}^\mathrm{det})^{5/6} f^{-7/6}$
where $D_\mathrm{L}$ is the luminosity distance to the source
and 
$\mathcal{M}^\mathrm{det} = (1+z) (m_1 m_2)^{3/5}/(m_1+m_2)^{1/5}$
is the detector-frame (redshifted) chirp mass, 
which gives the leading-order evolution of the binary amplitude and phase, 
and $z$ is the source redshift.

\begin{figure}[htbp]
  \begin{center}
 \begin{center}
    \includegraphics[keepaspectratio=true,height=70mm]{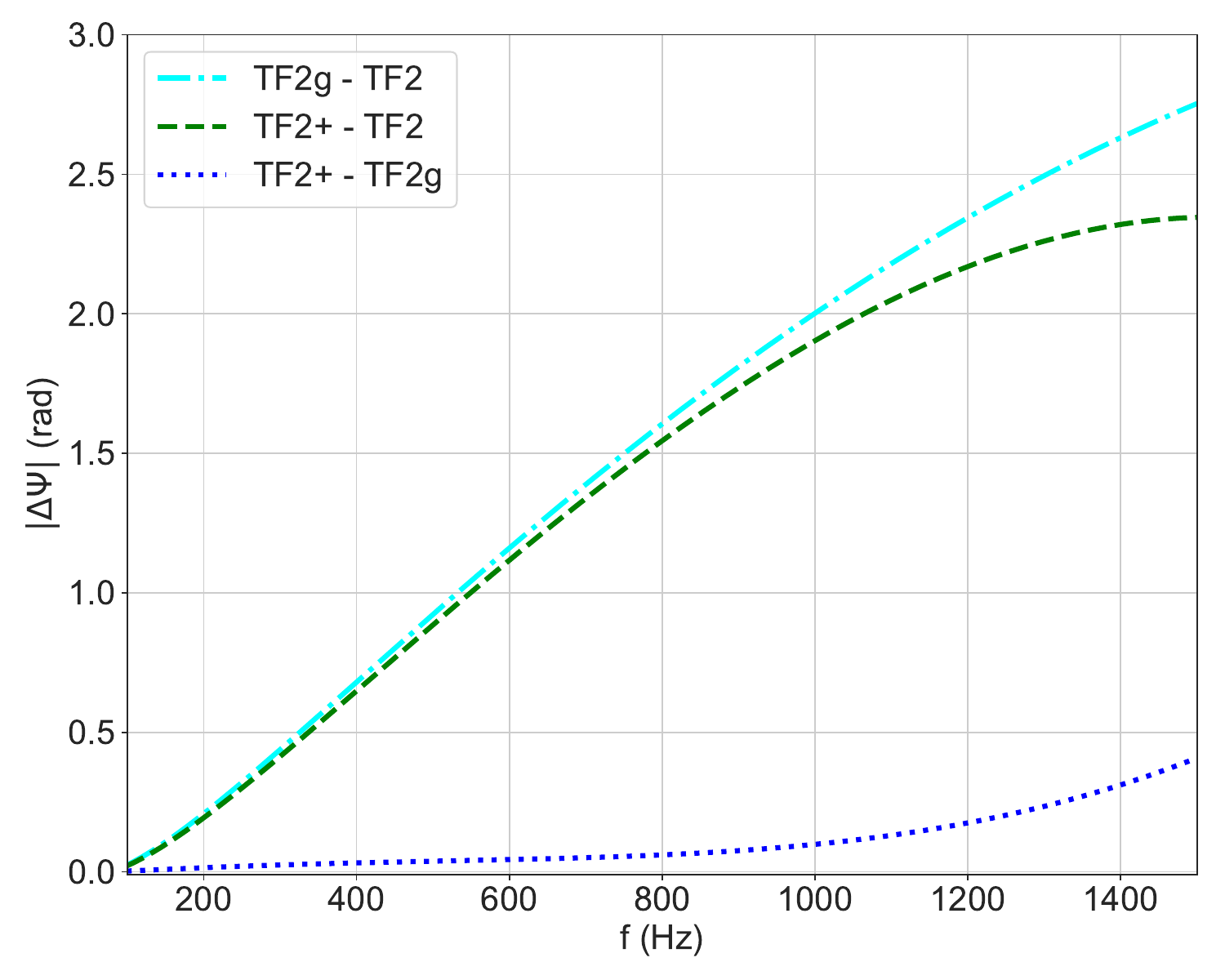}\\
 \end{center}
  \caption{
Absolute magnitude of the point-particle phase difference of the \texttt{TF2+} (green, dashed) and \texttt{TF2g} (cyan, dot-dashed) models from the \texttt{TF2} model
for the unequal-mass case of $(m_1,~m_2)=(1.68~\solM,~1.13~\solM)$.
Phase difference between the \texttt{TF2+} and \texttt{TF2g} models is also shown (blue, dotted), which is less than $0.5~(\mathrm{rad})$ up to $1500~\mathrm{Hz}$.
Here, we align different waveform models for $10~\mathrm{Hz}\leq f \leq100~\mathrm{Hz}$ to plot the phase difference.
}%
\label{fig:DiffPhase_PP}
\end{center}
\end{figure}

$\Psi_\mathrm{BBH}(f)$ consists of the point-particle and spin parts.
We use the \texttt{TF2} model~\cite{Dhurandhar:1992mw, Buonanno:2009zt, Blanchet:2013haa}
and extended models of the \texttt{TF2} model: the \texttt{TF2g} and \texttt{TF2+} models
as a point-particle part.
For the \texttt{TF2} model, the 3.5PN order formula is employed for the phase 
as summarized in Refs.~\cite{Buonanno:2009zt, Khan:2015jqa}.
For the \texttt{TF2g} model, the phase is extended to the quasi-5.5PN order, 
which is derived by the Taylor expansion of the EOB dynamics 
taking into account the notion in the test particle limit~\cite{Messina:2019uby}.
We set the uncalculated terms at 4PN order and beyond to zero.
For the \texttt{TF2+} model, both the phase and the amplitude are extended to the 6PN order, 
which is derived by the fitting to the \texttt{SEOBNRv2} model~\cite{Taracchini:2013rva} for the BNS mass range~\cite{Kawaguchi:2018gvj}.
The added higher PN order terms enable us to reduce the tidal deformability biasing.
In Fig.~\ref{fig:DiffPhase_PP}, we show absolute magnitude of the point-particle phase difference of the \texttt{TF2+} and \texttt{TF2g} models from the \texttt{TF2} model
for the unequal-mass case of $(m_1,~m_2)=(1.68~\solM,~1.13~\solM)$.
The phase difference between the \texttt{TF2+} and \texttt{TF2g} models is also shown, which is less than $0.5~(\mathrm{rad})$ up to $1500~\mathrm{Hz}$.
Here, we align different waveform models for $10~\mathrm{Hz}\leq f \leq100~\mathrm{Hz}$ to plot the phase difference.
All waveform models for BNSs used in our parameter estimation analyses
assume that the spins of component objects are aligned with the orbital angular momentum
and incorporate the 3.5PN order formula in couplings between the orbital angular momentum 
and the component spins~\cite{Bohe:2013cla},
3PN order formula in point-mass spin-spin, 
and self-spin interactions~\cite{Arun:2008kb, Mikoczi:2005dn} as summarized in Ref.~\cite{Khan:2015jqa}.
The different PN order terms for spin effects could help to break degeneracy between parameters.

We also use the PN formula for the tidal effects (\texttt{PNTidal}).
Recently, Ref.~\cite{Henry:2020ski} have derived the complete and correct form up to 2.5PN order (relative 5+2.5PN order)
for the mass quadrupole, current quadrupole, and mass octupole contributing to the GW tidal phase $\Psi_\mathrm{PNTidal}(f)$.
For the mass quadrupole interactions using data analysis
as a function of the dimensionless tidal deformability of each object $\Lambda_{1,2}$, we rewrite it as~\cite{Narikawa:2021pak}
\begin{widetext}
\begin{eqnarray}
 \Psi_\mathrm{PNTidal}(f) = &&\frac{3}{128\eta} x^{5/2} \sum_{A=1}^{2} \Lambda_A X_A^4 
 \left[ -24(12-11 X_A) - \frac{5}{28} (3179-919 X_A - 2286 X_A^2 + 260 X_A^3 ) x \right. \nonumber \\
 && +24 \pi (12 - 11 X_A) x^{3/2} \nonumber \\
 && -5 \left( \frac{193986935}{571536} - \frac{13060861}{381024} X_A - \frac{59203}{378} X_A^2
 - \frac{209495}{1512} X_A^3 + \frac{965}{54} X_A^4 - 4 X_A^5 \right) x^2 \nonumber \\
 &&  \left. + \frac{\pi}{28} (27719 - 22415 X_A + 7598 X_A^2 - 10520 X_A^3) x^{5/2}  \right], 
\label{eq:NUTTidal_phase}
\end{eqnarray}
\end{widetext}
where $x=[\pi M (1+z) f]^{2/3}$ is the dimensionless PN expansion parameter, 
and $\eta=m_1 m_2 / (m_1+m_2)^2$ is the symmetric mass ratio.
The tidal phase terms at 5+2PN order (line three) in Eq.~(\ref{eq:NUTTidal_phase}) corresponds to the complete version of uncalculated coefficients in Ref.~\cite{Henry:2020ski} and the 5+2.5PN terms (line four) in Eq.~(\ref{eq:NUTTidal_phase}) corresponds to the correct version in Ref.~\cite{Henry:2020ski}\footnote{Following Ref.~\cite{Henry:2020ski_v4}, Eq.~(\ref{eq:NUTTidal_phase}) has been modified in the latest version, which is described in the Erratum~\cite{Narikawa_Erratum}. Two coefficients of 7PN order have been modified.}.
The tidal amplitude $A_\mathrm{PNTidal}(f)$ is up to 5+1PN order as summarized in Refs.~\cite{Hotokezaka:2016bzh, Kawaguchi:2018gvj, Narikawa:2019xng}.

For realistic cases, the tidal contributions to the GW phase are dominated by the symmetric contribution in terms of  $\tilde{\Lambda}$.
Motivated by this fact, ignoring the asymmetric contributions in terms of $\delta\tilde{\Lambda}$, the $\tilde{\Lambda}$-form is obtained as\footnote{Eq.~(\ref{eq:NUTTidal_phase_LamtForm}) is obtained from Equation~(\ref{eq:NUTTidal_phase}) by replacing $X_A$ by $1/2$ and $\Lambda_A$ by $\tilde{\Lambda}$.}
\begin{eqnarray}
&& \Psi_\mathrm{PNTidal}^{\tilde{\Lambda}-\mathrm{form}} (f) = 
 \frac{3}{128\eta} \left( - \frac{39}{2} \tilde{\Lambda} \right) x^{5/2} \nonumber \\
 && \times \left[ 1 + \frac{3115}{1248} x - \pi x^{3/2} + \frac{29323235}{3429216} x^2 - \frac{2137}{546} \pi x^{5/2} \right]. \nonumber \\
\label{eq:NUTTidal_phase_LamtForm} 
\end{eqnarray}
The NR calibrated tidal waveform models, \texttt{KyotoTidal}~\cite{Kawaguchi:2018gvj}, \texttt{NRTidal}~\cite{Dietrich:2017aum}, and \texttt{NRTidalv2}~\cite{Dietrich:2019kaq}, are constructed by extension of this form\footnote{Similarly to the modification of Eq.~(\ref{eq:NUTTidal_phase}), Eq.~(\ref{eq:NUTTidal_phase_LamtForm}) has also been modified in the latest version. One coefficient of 7PN order has been modified.}.
Therefore, we use this form only when comparing the PN tidal with the NR calibrated waveform models in Sec.~
\ref{sec:results}. 
The \texttt{NRTidal} model is a NR simulation calibrated model, which provides a tidal description based on a fit of hybrid EOB-NR waveforms with the Pad\'{e} approximation.
The \texttt{NRTidalv2} model is an upgrade of the \texttt{NRTidal} model, by enforcing consistency with the analytical 5+2.5PN knowledge at the low frequency limit and including a tidal amplitude correction.
The \texttt{KyotoTidal} model is another NR calibrated model for the inspiral phase of BNS mergers employing hybrid waveforms combining \texttt{SEOBNRv2T}~\cite{Hinderer:2016eia, Steinhoff:2016rfi} and NR. 
It employs the 5+2.5PN order tidal-phase formula by multiplying the $\tilde{\Lambda}$ by a nonlinear correction to model the tidal part of the phase. It is calibrated only up to $1000~\mathrm{Hz}$ to avoid uncertainties 
in the postinspiral regime associated with physical effects that are not taken into the simulation.

\begin{figure}[htbp]
  \begin{center}
 \begin{center}
    \includegraphics[keepaspectratio=true,height=70mm]{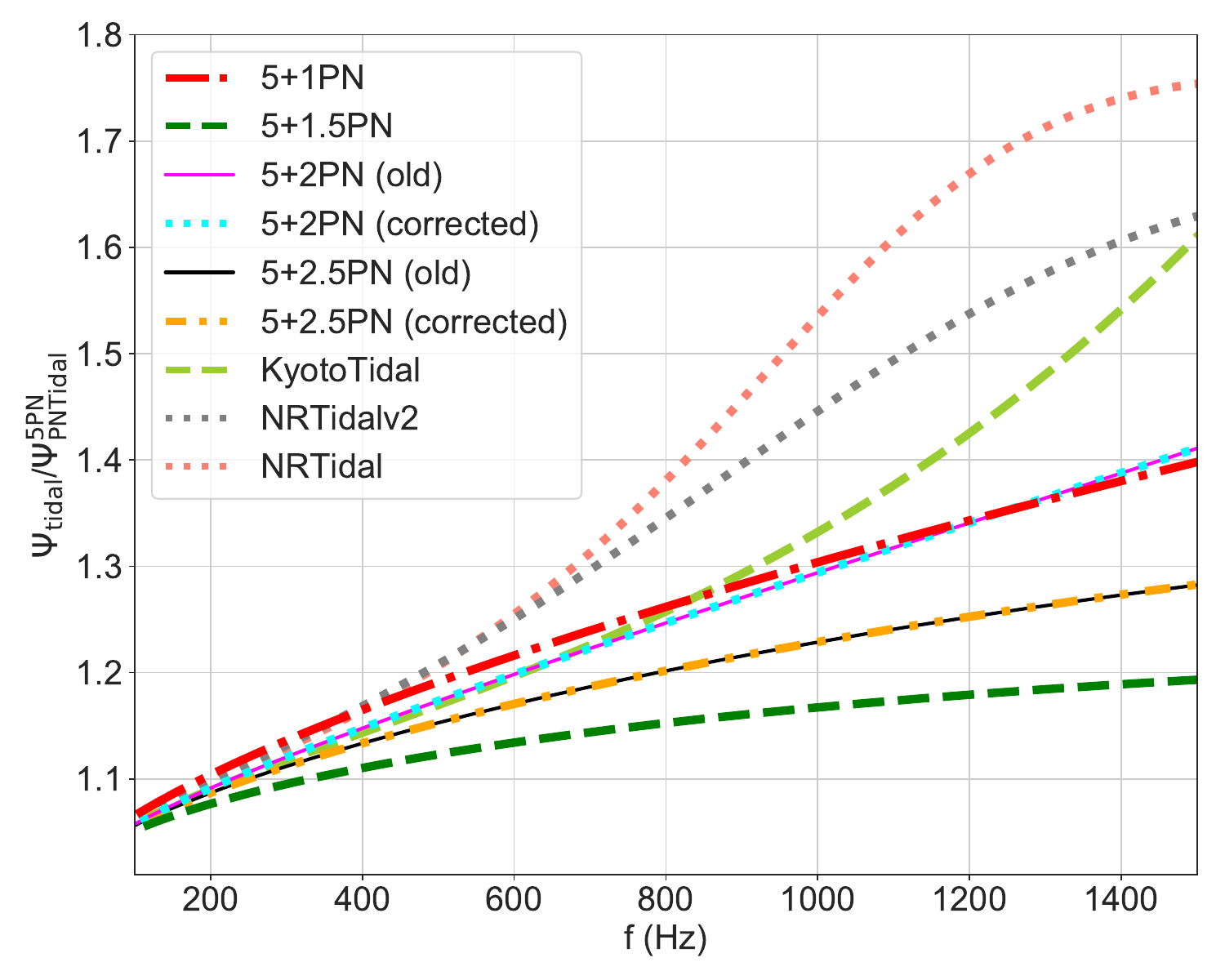}\\
 \end{center}
  \caption{
Tidal phase in the frequency domain normalized by the leading-order PN tidal phase at the relative 5PN
for the same unequal-mass case as used in Fig.~\ref{fig:DiffPhase_PP}.
We show 5+1PN (red, dot-dashed), 5+1.5PN (green, dashed), the old version of 5+2PN (magenta, solid), 
the corrected version of 5+2PN (cyan, dotted), the old version of 5+2.5PN (black, solid), 
and the corrected version of 5+2.5PN (orange, loosely dot-dashed) for \texttt{PNTidal}.
We also show the NR calibrated models: \texttt{KyotoTidal} (yellowgreen), \texttt{NRTidal} (salmon), and \texttt{NRTidalv2} (gray) for comparison.
}%
\label{fig:PhaseRatio_PN_Kyoto_NRTidal}
\end{center}
\end{figure}

\begin{figure}[htbp]
  \begin{center}
 \begin{center}
    \includegraphics[keepaspectratio=true,height=70mm]{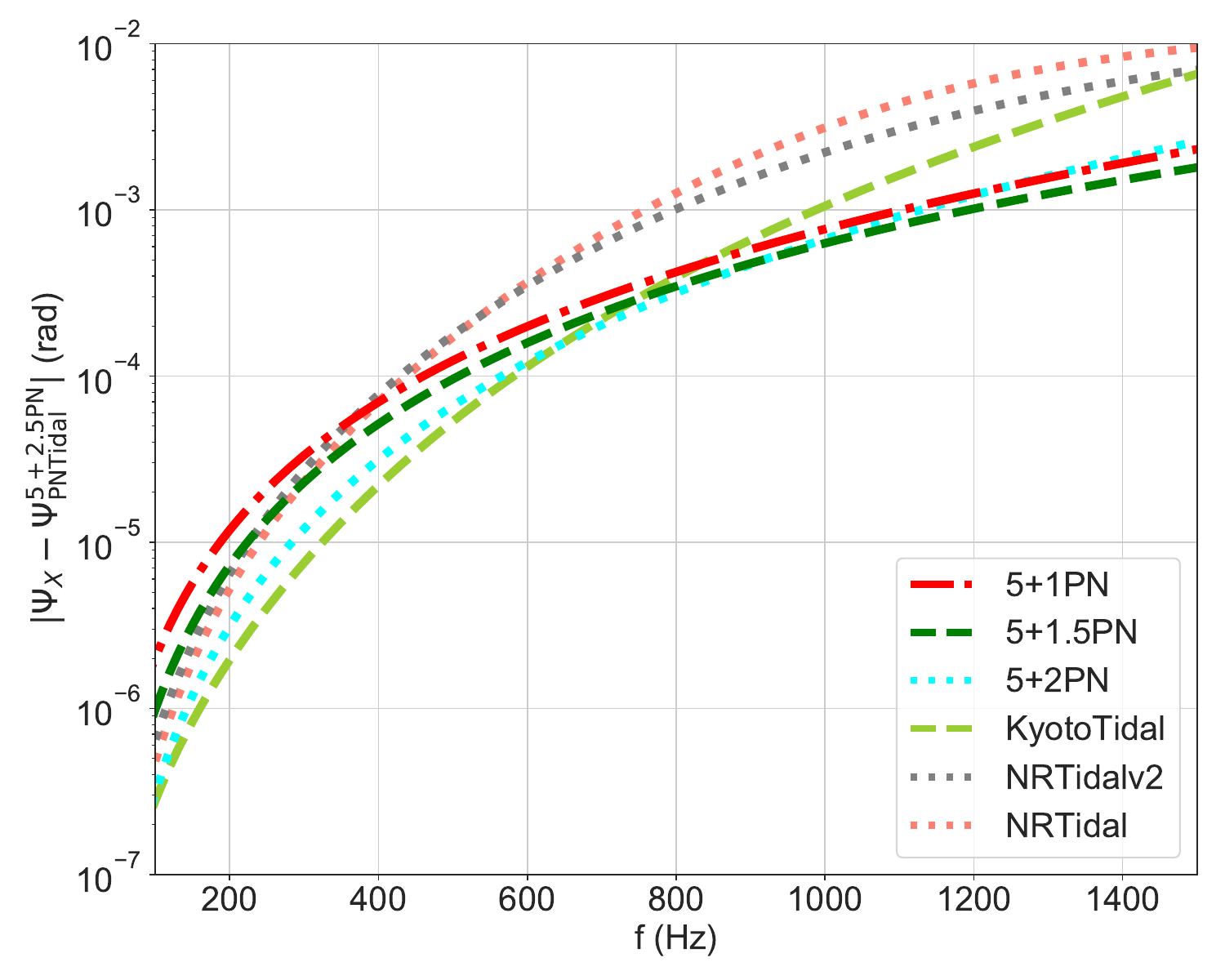}\\
 \end{center}
  \caption{
Absolute magnitude of the tidal phase difference of the different PN orders and NR calibrated models from the \texttt{PNTidal} model with 5+2.5PN order.
Here, they are calculated for the same unequal-mass case as used in Fig.~\ref{fig:DiffPhase_PP}.
}%
\label{fig:PhaseDifference_PN_Kyoto_NRTidal}
\end{center}
\end{figure}

In Fig.~\ref{fig:PhaseRatio_PN_Kyoto_NRTidal}, we present the tidal phase in the frequency domain of different PN order.
Here, they are divided by the leading-order tidal phase at the relative 5PN, $\Psi_\mathrm{PNTidal}^\mathrm{5PN}(f)$, and 
calculated for the same unequal-mass case as used in Fig.~\ref{fig:DiffPhase_PP}
and $(\Lambda_1,~\Lambda_2)=(102.08,~840.419)$, corresponding to ($\tilde{\Lambda},~\delta\tilde{\Lambda})=(292,~62.0784)$ motivated from the EOS APR4~\cite{Akmal:1998cf}.
We compare the old and corrected version of tidal phase in the 5+2PN and 5+2.5PN orders.
As an old model, we use Agathos {\it et al.} model~\cite{Agathos:2015uaa}. As a corrected model, we use Eq.~(\ref{eq:NUTTidal_phase}).
The phase difference between the old and corrected versions of the \texttt{PNTidal} model in the 5+2PN and 5+2.5PN orders are numerically very small,
which are at most $10^{-5}\tilde{\Lambda}$ (rad) up to $1500~\mathrm{Hz}$.
(See Fig.~\ref{fig:PhaseDifference_PNTidal} in Appendix \ref{sec:Old_vs_Corrected_PNTidal},
which shows the absolute magnitude of the phase difference between the old and corrected versions of the \texttt{PNTidal} model in the 5+2PN and 5+2.5PN orders.) 
We also show the NR calibrated models: the \texttt{KyotoTidal}, \texttt{NRTidal}, and \texttt{NRTidalv2} models, for comparison.
The phase shift does not monotonically change as the PN order in \texttt{PNTidal} increases.
Since the half-PN orders, 5+1.5PN and 5+2.5PN, are repulsive effect,
the terms at 5+1PN and 5+2PN orders give larger phase shift than the half-PN orders
and are closer to the \texttt{KyotoTidal} model up to around $1000~\mathrm{Hz}$ than the half-PN orders.
Other NR calibrated models, \texttt{NRTidalv2} and \texttt{NRTidal}, give larger phase shift as the frequency increases.
Figure~\ref{fig:PhaseDifference_PN_Kyoto_NRTidal} shows the absolute magnitude of the tidal phase difference of the different PN orders and NR calibrated models from the \texttt{PNTidal} model with 5+2.5PN order.

\begin{table}[htbp]
\begin{center}
\caption{
PN waveform models used to reanalyze GW170817 and GW190425.
The \texttt{TF2}, \texttt{TF2g}, and \texttt{TF2+} models are for a point-particle part.
The \texttt{TF2} model employs the 3.5PN order formula for the phase and 3PN order formula for the amplitude.
For the \texttt{TF2g} model, the phase is extended to the quasi-5.5PN order.
For the \texttt{TF2+} model is a phenomenological extended to 6PN order for both the phase and amplitude, which is derived by fitting to the \texttt{SEOBNRv2} model. 
We treat the same spin description for all models as follows:
aligned spins, 3.5PN order formula in spin-orbit interactions, 3PN order formula in spin-spin, and self-spin interactions for the phase.
The spin amplitude is up to 3PN order.
The \texttt{PNTidal} model employs the PN formula up to 5+2.5PN order for the phase and 5+1PN order for the amplitude.
}
\vspace{5pt}
\begin{tabular}{c|cc}
\hline \hline
Model name & Amplitude      & Phase \\ \hline
\texttt{TF2} & 3PN & 3.5PN \\ 
\texttt{TF2g} & 3PN & quasi-5.5PN \\ 
\texttt{TF2+} & 6PN & 6PN \\ \hline
\texttt{PNTidal} & 5+1PN & 5+2.5PN \\ 
\hline \hline
\end{tabular}
\label{table:WFs}
\end{center}
\end{table}

\subsection{Bayesian inference}
\label{sec:Bayesian}
We employ Bayesian inference for GW parameter estimation and model comparison 
between waveform models (see Refs.~\cite{Thrane:2018qnx, Christensen:2020, Veitch:2014wba} for review).
Given data $d$, which contains the signal and the noise,
according to Bayes's theorem, 
the posterior distribution of the signal parameters $\theta$ that the waveform $\tilde{h}(\theta)$ depends on
is given by 
\begin{eqnarray}
 p(\theta | d) = \frac{\mathcal{L}(d | \theta)\pi(\theta)}{\mathcal{Z}},
\end{eqnarray}
where $\mathcal{L}(d | \theta)$ is the likelihood function of the data for given parameters $\theta$, 
$\pi(\theta)$ is the prior distribution for $\theta$,
and $\mathcal{Z}$ is the evidence.
By assuming stationarity and Gaussianity for the detector noise,
the likelihood function is evaluated as,
\begin{eqnarray}
 \mathcal{L}(d | \theta) \propto \exp \left[ - \frac{\langle d - h(\theta) | d - h(\theta) \rangle}{2}\right],
\end{eqnarray}
where the noise-weighted inner product $\langle \cdot | \cdot \rangle$ is defined by
\begin{eqnarray}
 \langle a | b \rangle := 4 \mathrm{Re} \int_{f_\mathrm{low}}^{f_\mathrm{high}} df \frac{\tilde{a}^*(f) \tilde{b}(f)}{S_n(f)},
\end{eqnarray}
using the noise power spectrum density $S_n(f)$. 
We use $S_n(f)$ obtained with the \texttt{BayesLine}~algorithm~\cite{Cornish:2014kda, Littenberg:2015kpb, Chatziioannou:2019zvs}, which model the PSD as a cubic spline for the broadband structure and a sum of Lorentzians for the line noise.
The lower limit of the integration $f_\mathrm{low}$ is the seismic cutoff frequency
and the higher limit $f_\mathrm{high}$ is the cutoff frequency of waveforms.
Similarly to the previous analyses, we set the lower frequency cutoff $f_\mathrm{low}=23~\mathrm{Hz}$ for GW170817~\cite{Abbott:2018wiz} and $19.4~\mathrm{Hz}$ for GW190425~\cite{Abbott:2020uma}.
As a frequency range becomes higher, tidal information becomes richer, while postinspiral physics becomes more important.
PN inspiral models are less valid due to the lack of postinspiral effects in the high frequency range.
To restrict the analysis to the inspiral regime of the signals, we conservatively set the upper frequency cutoff $f_\mathrm{high}=1000~\mathrm{Hz}$, which agrees with the calibration range of the KyotoTidal model.

The evidence is obtained as the likelihood marginalized over the prior volume,
\begin{eqnarray}
 \mathcal{Z} = \int d\theta \mathcal{L}(d | \theta) \pi(\theta).
\end{eqnarray}
To perform model comparison between the waveform models A and B,
we compute the ratio between two different evidences, called the Bayes factor\footnote{$\log_{10}\mathrm{BF} > 1.5$ is often interpreted as a strong preference for one model over another, 
and $\log_{10}\mathrm{BF} > 2$ is interpreted as decisive evidence~\cite{Jeffreys:1961}.},
\begin{eqnarray}
 \mathrm{BF}_\mathrm{A/B} = \frac{\mathcal{Z}_\mathrm{A}}{\mathcal{Z}_\mathrm{B}}.
\end{eqnarray}
The one-dimensional posterior for a specific parameter is obtained 
by marginalizing the multidimensional posterior over the other parameters.
We compute posterior probability distribution functions (PDFs)
by using Bayesian stochastic sampling based on the nested sampling algorithm~\cite{Skilling:2006,Veitch:2009hd}.
Specifically, we use the parameter estimation software, LALInference~\cite{Veitch:2014wba},
which is one of the software programs of LIGO Algorithm Library (LAL) software suite~\cite{LAL}. 
There are 2048 live points for the nested sampling algorithm and a tolerance equal to 0.1. About $2\times10^4$ samples from each analysis are collected to create posteriors.

\subsection{Source parameters and their priors}
\label{sec:parameters}
The source parameters and their prior probability distributions are basically chosen to follow 
those adopted in the papers on LVC's analyses of GW170817~\cite{Abbott:2018wiz},
GW190425~\cite{Abbott:2020uma}, and
our analyses of them~\cite{Narikawa:2018yzt, Narikawa:2019xng}.
We mention the specific choices adopted below.

For BNS coalescences, the parameters are 
the component masses $m_{1,2}$, where we assume $m_1\geq m_2$;
the orbit-aligned dimensionless spin components of the stars $\chi_{1z,2z}$,
defined as 
$\chi_{1z,2z} = S_{1z,2z} / (m_{1,2}^2)$ 
where $S_{1z,2z}$ are the magnitudes of the spin angular momenta of the components;
the binary tidal deformability $\tilde{\Lambda}$; 
the asymmetric contribution of tidal deformability $\delta\tilde{\Lambda}$;
the luminosity distance to the source $D_\mathrm{L}$; 
the binary inclination angle $\theta_\mathrm{JN}$
defined as $\cos\theta_\mathrm{JN}=\hat{\boldsymbol{J}} \cdot \hat{\boldsymbol{N}}$, 
which is the angle between the total angular momentum $\boldsymbol{J}$ and the line of sight $\boldsymbol{N}$
(a hat symbol $\hat{}$ represents a unit vector); 
and the polarization angle is $\psi$.
In our analyses, we marginalized over the coalescence time $t_c$ and the phase at the coalescence time $\phi_c$ 
semianalytically.
For GW170817, we fix the sky location to the position of AT 2017gfo, which is an electromagnetic counterpart of GW170817~\cite{Soares-Santos:2017lru, LIGOScientific:2017ync, J-GEM:2017tyx}.

We employ a uniform prior on the detector-frame component 
masses $m_{1,2}^\mathrm{det}$ in the range $[0.5,~5.0]M_\odot$
with an additional constraint on the mass ratio $q=m_2/m_1>0.0556$,
and the detector-frame total mass $M^\mathrm{det}$ in the range $[1.0,~10.0]M_\odot$.
We assume a uniform prior on the spin magnitudes $\chi_{1z,2z}$ in the range $[-0.05,~0.05]$\footnote{The low-spin prior for the LVC's analyses is a uniform on the magnitudes of the spins  in the range $[-0.05,~0.05]$~\cite{TheLIGOScientific:2017qsa, Abbott:2018wiz}.},
which is astrophysically motivated from the maximum spin observed in Galactic BNSs that will merge within the Hubble time~\cite{Burgay:2003jj}.
We assume a uniform prior both on the binary tidal deformability $\tilde{\Lambda}$ in the range $[0,~3000]$
and the asymmetric contribution $\delta\tilde{\Lambda}$ in the range $[-3000,~3000]$.
On the other hand, in the LVC's analyses, they analyze with a uniform prior on the component $\Lambda_{1,2}$, 
and weight the posterior by dividing by the prior to effectively obtain a uniform prior on $\tilde{\Lambda}$~\cite{TheLIGOScientific:2017qsa, Abbott:2018wiz}.
We note that the asymmetric contribution $\delta\tilde{\Lambda}$ terms are ignored
only for the $\tilde{\Lambda}$-form defined as Eq.~(\ref{eq:NUTTidal_phase_LamtForm}) and the NR calibrated models: the \texttt{KyotoTidal}, \texttt{NRTidal}, and \texttt{NRTidalv2} models.
This ignorance reduces the prior volume.

For GW190425, we estimate the sky location of the source with an isotropic prior.
We assume the priors for the detector-frame component mass, the spins, and the binary tidal deformability 
are the same as the ones for GW170817.

In our analyses, we use the EOS-insensitive relations between tidal deformability and spin-induced quadrupole moment called Love-Q relations predicted 
by theoretical studies~\cite{Yagi:2013bca, Yagi:2013awa}, which have been used in Refs.~\cite{Abbott:2018wiz, Abbott:2018exr, Abbott:2020uma, Narikawa:2018yzt, Narikawa:2019xng}.

\section{Results}
\label{sec:results}
We reanalyze the data around BNS coalescence events GW170817 and GW190425 
using the corrected \texttt{PNTidal} model\footnote{All results shown in this paper have been obtained with the previous expression of the tidal phase, which is described in the previous version of this paper. The modification to the tidal phase Eqs.~(\ref{eq:NUTTidal_phase}) and (\ref{eq:NUTTidal_phase_LamtForm})
is numerically small. The phase difference between the previous and modified versions of the tidal phase Eq.~(\ref{eq:NUTTidal_phase}) is less than $\mathcal{O}(10^{-3})$ (rad) up to 1000~Hz for a binary neutron star with $m_A=1.68~M_\odot$, $m_B=1.13~M_\odot$, $\Lambda_A=102$, and $\Lambda_B=840$.
The difference is less than $\mathcal{O}(10^{-2})$ (rad) up to 200 Hz for a binary exotic compact object with $m_A=13~M_\odot$, $m_B=6.6~M_\odot$, $\Lambda_A=-700$, and $\Lambda_B=-1400$.
Therefore, their impact on results in this paper is negligible.}.
We use the publicly abailable data on the Gravitational Wave Open Science Center
(https://www.gw-openscience.org) released by the LIGO-Virgo-KAGRA (LVK) Collaborations~\cite{LIGOScientific:2019lzm}.

As a sanity check, we compare the results for GW170817 and GW190425 by using the old and corrected versions of the \texttt{PNTidal} model for $|\chi_{1z,2z}| \leq 0.05$ and the upper frequency cutoff $f_\mathrm{high}=1000~\mathrm{Hz}$.
We show the results for representative model, the \texttt{TF2+\_PNTidal} model.
We find good agreement between the marginalized symmetric 90\% credible intervals of the binary tidal deformability $\tilde{\Lambda}$ 
by using the old and corrected versions of the \texttt{PNTidal} model as expected by the tidal phase difference shown in Fig.~\ref{fig:PhaseDifference_PN_Kyoto_NRTidal} as follows:
$579^{+486}_{-437}$ with the old model and $574^{+485}_{-425}$ with the corrected model for GW170817,
and 
$297^{+589}_{-266}$ with the old model and $295^{+578}_{-265}$ with the corrected model for GW190425.
The log Bayes factor between the old and corrected model, $\log\mathrm{BF}_\mathrm{corrected/old}=0.25$ for GW170817 and $-0.15$ for GW190425, which shows no preference for the corrected model against the old model.

The 90\% credible intervals of the source parameters for the \texttt{TF2+\_PNTidal} model
are presented in Table~\ref{table:all_GW170817_GW190425_TF2plus_PNTidal}.
We show the estimates of 
mass parameters (the component masses $m_{1,2}$, 
the chirp mass $\mathcal{M}$, the detector-frame chirp mass $\mathcal{M}^\mathrm{det}$, 
the mass ratio $q$, and the total mass $M$);
the effective inspiral spin parameter $\chi_\mathrm{eff} = (m_1 \chi_{1z} + m_2 \chi_{2z})/M$, 
which is the most measurable combination of spin components~\cite{Ajith:2009bn, Santamaria:2010yb};
the luminosity distance to the source $D_\mathrm{L}$;  
and the binary tidal deformability $\tilde{\Lambda}$. 
Since we find that the asymmetric contribution $\delta\tilde{\Lambda}$ is uninformative relative to a prior for both events, we do not show them.
The source-frame masses are derived 
by assuming a value of the Hubble constant $H_0 = 69~\mathrm{km}~\mathrm{s}^{-1}~\mathrm{Mpc}^{-1}$ 
(adopted from Planck 2013 results~\cite{Ade:2013sjv}).
The posterior PDFs of parameters obtained with the \texttt{TF2+\_PNTidal} model are presented in Appendix \ref{sec:WFsys}.
We check that estimates of all parameters we obtained by using the corrected version of the \texttt{TF2+\_PNTidal} model
are broadly consistent with the previous results by using the old version of the \texttt{TF2+\_PNTidal} model  presented in Refs.~\cite{Narikawa:2019xng}.

First, we discuss the waveform systematics with respect to the point-particle part 
in the inference of $\tilde{\Lambda}$.
Figure~\ref{fig:Lamt_GW170817_GW190425_PPs} shows 
the posterior PDFs of $\tilde{\Lambda}$ for GW170817 (left panel) and GW190425 (right panel) 
estimated by using three different point-particle part models, \texttt{TF2}, \texttt{TF2g}, and \texttt{TF2+}, 
employing the same tidal part \texttt{PNTidal}.
The estimates of $\tilde{\Lambda}$ show the absence of significant systematic difference 
associated with the phase difference among waveform models for point-particle part.
This result shows that the higher-order point-particle terms, which are incorporated in \texttt{TF2g} and \texttt{TF2+}, do not significantly affect the estimate of $\tilde{\Lambda}$ for GW170817 and GW190425 for $f_\mathrm{high}=1000~\mathrm{Hz}$.
In Table \ref{table:logBF_GW170817_GW190425_PPs}, we present the log Bayes factor of the \texttt{TF2+\_PNTidal} model relative to the other point-particle baseline models: the \texttt{TF2g\_PNTidal} and \texttt{TF2\_PNTidal} models, 
$\log\mathrm{BF}_{\texttt{TF2+\_PNTidal}/\texttt{TF2g\_PNTidal},~\texttt{TF2\_PNTidal}}$,
for GW170817 and GW190425. 
The values indicate no preference among the different point-particle models by relying on the GW170817 and GW190425.
In Appendix~\ref{sec:WFsys},
we demonstrate the absence of significant systematic difference in source parameters
associated with the phase difference among point-particle baseline models.
Our inspiral-only analyses with the upper frequency cutoff $f_\mathrm{high}=1000~\mathrm{Hz}$ give larger $\tilde{\Lambda}$ and wider posterior distributions than the full analyses with $f_\mathrm{high}=2048~\mathrm{Hz}$ in Ref.~\cite{Abbott:2018wiz}, as discussed in Refs.~\cite{Dai:2018dca, Narikawa:2019xng, Gamba:2020wgg}.
The differences in the prior on the tidal deformability do not significantly affect the estimates of $\tilde{\Lambda}$ for GW170817, as shown in Refs.~\cite{De:2018uhw, Narikawa:2019xng}.
In order to compare with the results presented in Ref. \cite{Breschi:2021wzr},
we also reanalyze GW170817 with \texttt{TF2g\_PNTidal} template employing with high-spins prior 
($|\chi| \leq 0.89$). 
The posterior distribution for the $\tilde{\Lambda}$ shows 
$533_{-436}^{+542}$ at the 90\% symmetric credible region and $533_{-495}^{+445}$ at the 90\% highest-probability-density (HPD) credible region.
Our marginalized distribution and the 90\% credible regions in the $\tilde{\Lambda}$ are largely consistent with the posterior distribution for $f_\mathrm{max}=1~\mathrm{kHz}$ presented in Fig.~17 in Sec.~VII B of Ref. \cite{Breschi:2021wzr}.

Then, we examine systematic difference in the estimates of $\tilde{\Lambda}$ among different PN orders in the tidal phase for the \texttt{PNTidal} model.
Figure~\ref{fig:Lamt_GW170817_GW190425_differentPN_Kyoto_NRTidal} shows 
the posterior PDFs of $\tilde{\Lambda}$ for GW170817 (left panel) and GW190425 (right panel)
estimated with the different PN orders, from 5PN through 5+2.5PN, in the \texttt{PNTidal} model.
Here the point-particle baseline is the \texttt{TF2+} model in common.
The estimates of $\tilde{\Lambda}$ slightly depend on the PN orders in the tidal phase.
For both events, we find that the results are consistent with the phase shift as shown in Fig.~\ref{fig:PhaseRatio_PN_Kyoto_NRTidal} and an increase of the PN order does not lead to a monotonic change of estimates of $\tilde{\Lambda}$.
Since the leading-order terms at 5PN do not model the tidal effects efficiently at high frequency, the posterior distribution of $\tilde{\Lambda}$ is the  widest.
The inferred median values with 90\% symmetric and HPD credible regions of the marginalized posterior information of $\tilde{\Lambda}$ for GW170817 and GW190425 are summarized 
in Table \ref{table:Lamt_GW170817_GW190425_TF2plus_differentPN}.
As shown in Table \ref{table:Lamt_GW170817_GW190425_TF2plus_differentPN},
the terms up to 5+1PN and 5+2PN give slightly smaller values of median for $\tilde{\Lambda}$ than the terms at 5+1.5PN and 5+2.5PN.
These are related to the half-PN orders at 5+1.5PN and 5+2.5PN being repulsive effect. 
This PN order dependence is consistent with the GW170817-like signal injection studies in Ref.~\cite{Samajdar:2018dcx}.
In Table \ref{table:logBF_GW170817_GW190425_TF2plus_differentPN}, we show the log Bayes factor of the terms at 5+2.5PN relative to the other PN orders, $\log \mathrm{BF}_\mathrm{5+2.5PN/different~PN}$.
The values indicate no preference among the different PN orders over the 5+2.5PN terms by relying on the GW170817 and GW190425.

For comparison, we also present the posterior PDFs of $\tilde{\Lambda}$ estimated by the NR calibrated models,
the \texttt{KyotoTidal}, \texttt{NRTidalv2}, and \texttt{NRTidal} in Fig.~\ref{fig:Lamt_GW170817_GW190425_differentPN_Kyoto_NRTidal} and Table \ref{table:Lamt_GW170817_GW190425_TF2plus_Kyoto_NRTidal}.
These results are obtained by reanalyses with the same setting as the ones for the \texttt{TF2+\_PNTidal} model (we set the \texttt{TF2+} model as a point-particle part, the spin amplitude formula up to 3PN order, $|\chi_{1z,2z}| \leq 0.05$, and $f_\mathrm{high}=1000~\mathrm{Hz}$ as an upper frequency cutoff).
Figure \ref{fig:Lamt_GW170817_GW190425_differentPN_Kyoto_NRTidal} shows that 
posterior uncertainties of $\tilde{\Lambda}$ for the NR calibrated models are narrower than the PNTidal models.
It means that NR information improves inference of the tidal deformability.
The estimated values of $\tilde{\Lambda}$ reported in Table \ref{table:Lamt_GW170817_GW190425_TF2plus_Kyoto_NRTidal} quantitatively supports the fact.
The posterior PDFs of $\tilde{\Lambda}$ estimated by the terms at 5+1PN and 5+2PN are closer to the \texttt{KyotoTidal} model 
than the half-PN orders: 5+1.5PN and 5+2.5PN.
The \texttt{NRTidalv2} and \texttt{NRTidal} models give slightly smaller $\tilde{\Lambda}$ than the \texttt{KyotoTidal}.
These results are expected by the tidal phase shift shown in Fig.~\ref{fig:PhaseRatio_PN_Kyoto_NRTidal}.
In Table \ref{table:logBF_GW170817_GW190425_TF2plus_Kyoto_NRTidal}, we show the log Bayes factor of the \texttt{PNTidal} model
relative to the NR calibrated models, 
$\log\mathrm{BF}_{\texttt{PNTidal}/\mathrm{NR~calibrated~model}}$, 
for GW170817 and GW190425.
We also show the results for the \texttt{TF2+} model as a BBH (nontidal) hypothesis in the last row.
The values indicate no strong preference among the NR calibrated models 
over the \texttt{PNTidal} 
by relying on the GW170817 and GW190425.
We note that these NR calibrated models are constructed by extension from the $\tilde{\Lambda}$-form Eq.~(\ref{eq:NUTTidal_phase_LamtForm}) and the asymmetric contributions in terms of $\delta\tilde{\Lambda}$ are not included.
Therefore, we use the $\tilde{\Lambda}$-form defined as Eq.~(\ref{eq:NUTTidal_phase_LamtForm}) 
for calculating the Bayes factor here to fairly compare the \texttt{PNTidal} model with the NR calibrated waveform models. 
We find that the Bayes factor of the $\Lambda_{1,2}$-form Eq.~(\ref{eq:NUTTidal_phase}) is relative to the $\tilde{\Lambda}$-form Eq.~(\ref{eq:NUTTidal_phase_LamtForm}): 
$\log\mathrm{BF}_\mathrm{Eq.(3)/Eq.(4)}=-8.17$ for GW170817 and -4.11 for GW190425, 
which indicate that neglecting the $\delta\tilde{\Lambda}$ terms strongly affects the log Bayes factor.
This is related that ignoring the asymmetric contribution $\delta\tilde{\Lambda}$ terms reduces the prior volume.

Finally, we discuss constraints on EOS models for NSs by combining information obtained from GW170817 and GW190425 by using the \texttt{TF2+\_PNTidal} model.
In Fig.~\ref{fig:McLamt_RcMc_GW170817_GW190425}
we plot prediction of various EOS models for NSs on posterior of 
the binary tidal deformability and the chirp mass $\tilde{\Lambda}$-$\mathcal{M}$ (left panel)
as well as the chirp mass and chirp radius 
$\mathcal{M}$-$\mathcal{R}$ (right panel).
Here, the chirp radius $\mathcal{R}$ is a conveniently scaled dimensionful radius like parameter defined by $\mathcal{R}=2\mathcal{M}\tilde{\Lambda}^{1/5}$~\cite{Wade:2014vqa}.
Since the chirp mass - chirp radius measurement of a BNS system is an analogue of a mass - radius measurement of a single NS, it is useful to understand the constraints on the EOS of NSs.
The contours show 50\% and 90\% credible regions 
for GW170817 (red) and GW190425 (blue) estimated by using the \texttt{TF2+\_PNTidal} model.
The dots are the values predicted by various EOS models for NSs,
where we select binaries with equal masses.
Our constraints on EOS models for NSs by using the \texttt{TF2+\_PNTidal} model show that 
less compact models, MS1, MS1B, and H4, lie outside the 90\% credible region of $\tilde{\Lambda}$-$\mathcal{M}$ plane for GW170817 and they are disfavored,
which is consistent with the previous results obtained by using several waveform models for $f_\mathrm{high}=2048~\mathrm{Hz}$ (or $1000~\mathrm{Hz}$) presented in Refs.
~\cite{TheLIGOScientific:2017qsa, Abbott:2018wiz, Abbott:2018exr, LIGOScientific:2019eut, Landry:2018prl, Landry:2020vaw, Capano:2019eae, Narikawa:2019xng}.
The 90\% symmetric (HPD) allowed range of the chirp radius is 
$\mathcal{R} = 12.5_{-3.0}^{+1.6}~\mathrm{km}$
($\mathcal{R} = 12.5_{-2.4}^{+2.0}~\mathrm{km}$)
for GW170817 with the \texttt{TF2+\_PNTidal} waveform model, which is slightly weaker constraint than the constraint obtained with the \texttt{TF2+\_KyotoTidal} model~\cite{Narikawa:2019xng}.

\begin{widetext}

\begin{table}[htbp]
\begin{center}
\caption{
The 90\% credible intervals of source parameters for GW170817 and GW190425 
estimated using the \texttt{TF2+\_PNTidal} model
for $|\chi_{1z,2z}| \leq 0.05$ and the upper frequency cutoff $f_\mathrm{high}=1000~\mathrm{Hz}$.
(In Appendix \ref{sec:WFsys}, we demonstrate these results are robust to systematic uncertainty among three PN point-particle models.)
Mass values are quoted in the source frame, accounting for uncertainty in the source redshift
by assuming a value of the Hubble constant $H_0 = 69~\mathrm{km}~\mathrm{s}^{-1}~\mathrm{Mpc}^{-1}$. 
We give the 0\%-90\% interval for $m_1$, 
while we give the 10\%-100\% interval for $m_2$ and $q$.
We give symmetric 90\% credible intervals, i.e., 5\%--95\%, for the other parameters 
with the median as a representative value. 
We also give HPD interval for $\tilde{\Lambda}$.
(Since we find that the asymmetric contribution $\delta\tilde{\Lambda}$ is uninformative relative to a prior for both events, we do not show them.)
}
\vspace{5pt}
\begin{tabular}{lccc}
\hline \hline
Parameters & ~~GW170817 &  ~~GW190425 \\ \hline
Primary mass $m_1$ & $1.36-1.57~\solM$ & $1.62-1.89~\solM$ \\   
Secondary mass $m_2$ & $1.19-1.37~\solM$ & $1.44-1.68~\solM$ \\
Chirp mass $\mathcal{M}$ & $1.187^{+0.004}_{-0.002}~\solM$ & $1.436^{+0.022}_{-0.020}~\solM$ \\
Detector-frame chirp mass $\mathcal{M}^\mathrm{det}$ & $1.1976^{+0.0001}_{-0.0001}~\solM$ & $1.4867^{+0.0003}_{-0.0003}~\solM$ \\
Mass ratio $q=m_2/m_1$ & $0.76-1.00$ & $0.76-1.00$ \\
Total mass $M=m_1+m_2$ & $2.73^{+0.04}_{-0.01}~\solM$ & $3.31^{+0.06}_{-0.05}~\solM$ \\
Effective inspiral spin $\chi_\mathrm{eff}$ & $0.003^{+0.014}_{-0.008}$ & $0.009^{+0.015}_{-0.012}$ \\
Luminosity distance $D_\mathrm{L}$ & $40.2^{+7.0}_{-14.0}$~Mpc & $160^{+67}_{-73}$~Mpc\\
Binary tidal deformability $\tilde{\Lambda}$ (symmetric/HPD) & $574^{+485}_{-425}$ / $574^{+433}_{-467}$ & $295^{+578}_{-265}$ / $295^{+423}_{-295}$ \\
\hline \hline
\end{tabular}
\label{table:all_GW170817_GW190425_TF2plus_PNTidal}
\end{center}
\end{table}

\begin{figure*}[htbp]
  \begin{center}
\begin{tabular}{cc}
 \begin{minipage}[b]{0.45\linewidth}
 \begin{center}
    \includegraphics[keepaspectratio=true,height=80mm]{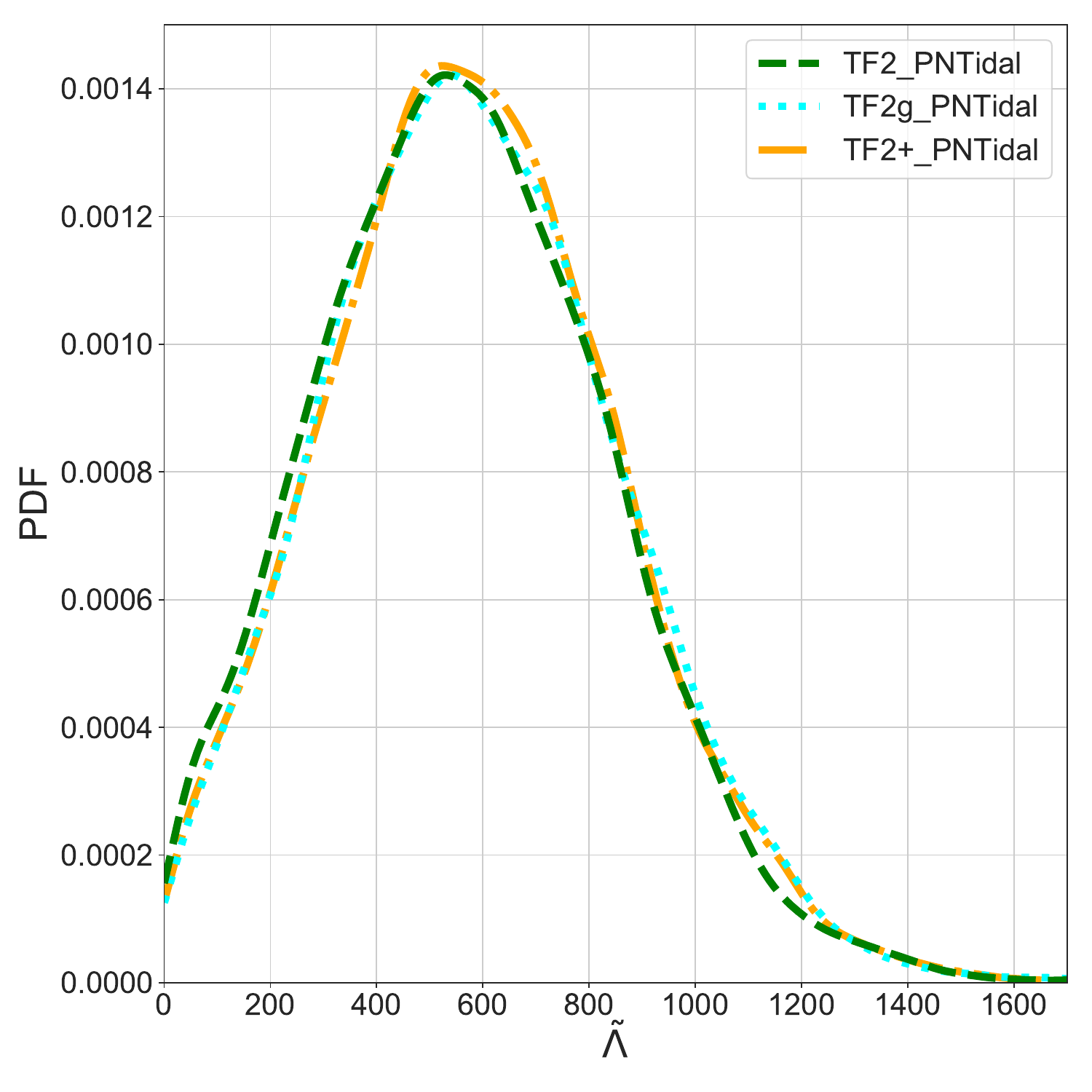}\\
 \end{center}
 \end{minipage}
 \begin{minipage}[b]{0.45\linewidth}
  \begin{center}
    \includegraphics[keepaspectratio=true,height=80mm]{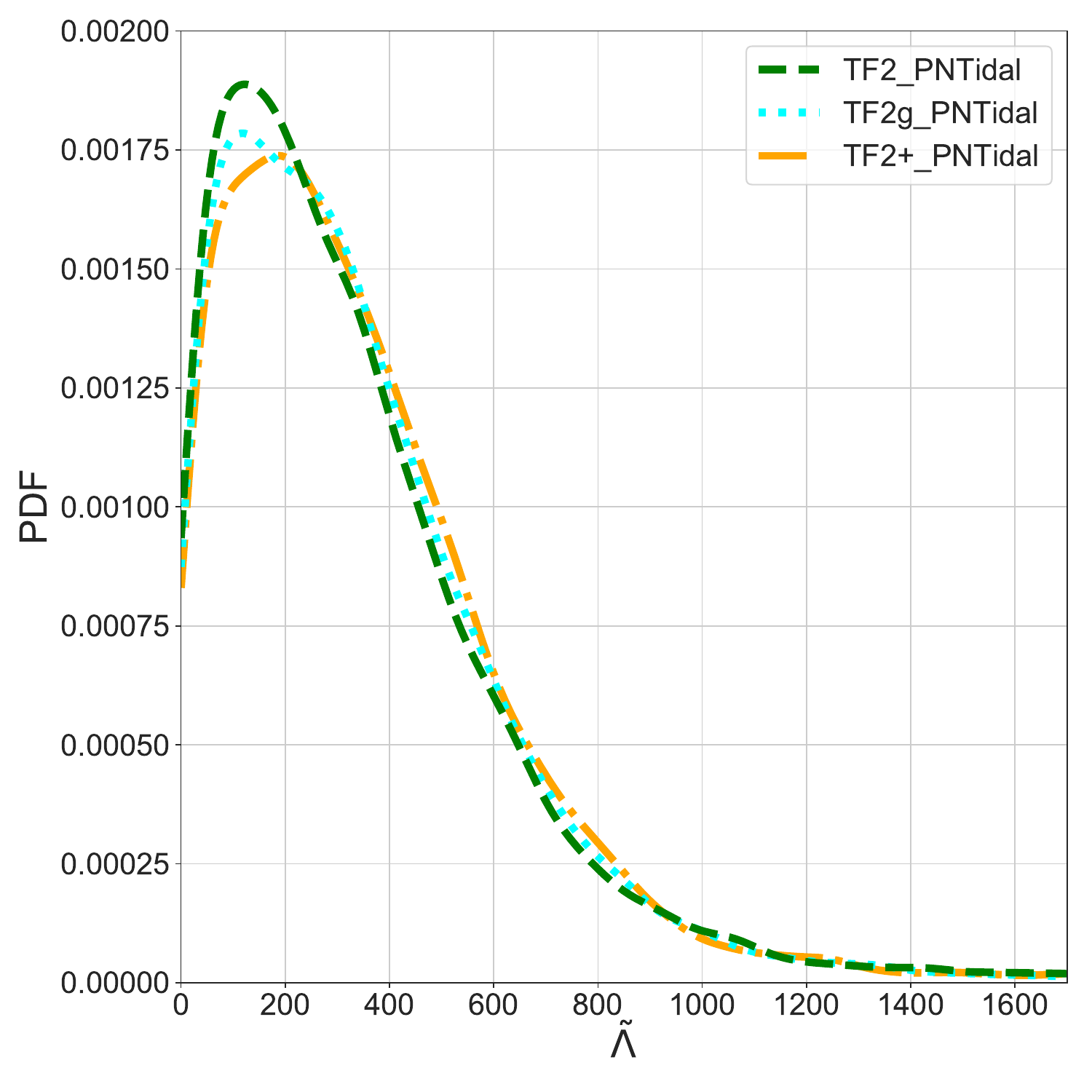}\\
 \end{center}
 \end{minipage}
\end{tabular}
  \caption{
Marginalized posterior PDFs of the binary tidal deformability $\tilde{\Lambda}$ 
for GW170817 (left) and GW190425 (right) 
estimated by using three point-particle part models, employing the same tidal model \texttt{PNTidal}
for $|\chi_{1z,2z}| \leq 0.05$ and the upper frequency cutoff $f_\mathrm{high}=1000~\mathrm{Hz}$.
The curves correspond to 
the \texttt{TF2\_PNTidal} (green), \texttt{TF2g\_PNTidal} (cyan), and \texttt{TF2+\_PNTidal} (orange) models.
The corresponding 90\% credible intervals are presented in Table \ref{table:all_GW170817_GW190425_TF2plus_PNTidal}.
There are no large difference in the estimates of $\tilde{\Lambda}$ among three PN point-particle models.
}
\label{fig:Lamt_GW170817_GW190425_PPs}
\end{center}
\end{figure*}

\begin{figure*}[htbp]
  \begin{center}
\begin{tabular}{cc}
 \begin{minipage}[b]{0.45\linewidth}
 \begin{center}
    \includegraphics[keepaspectratio=true,height=80mm]{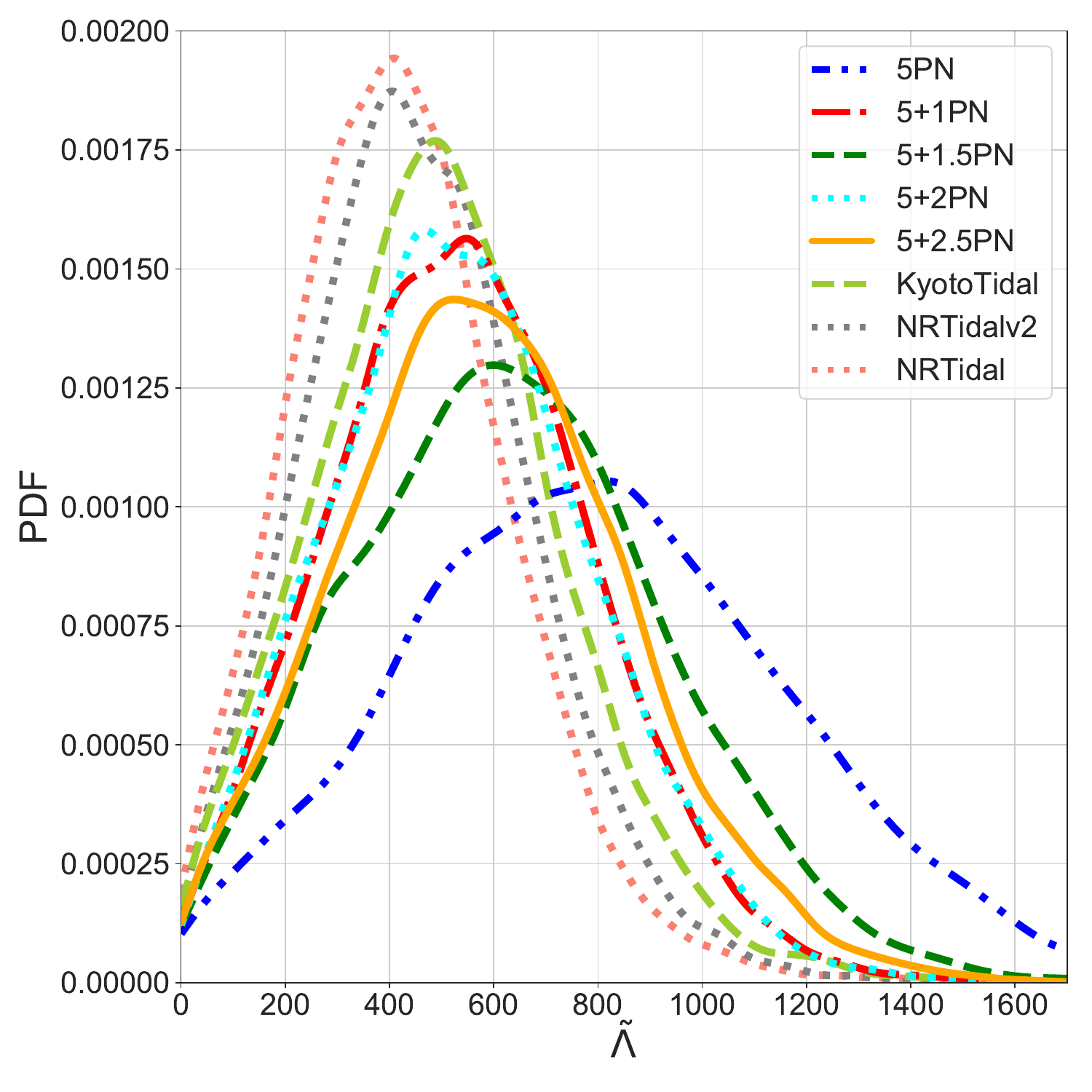}\\
 \end{center}
 \end{minipage}
 \begin{minipage}[b]{0.45\linewidth}
  \begin{center}
    \includegraphics[keepaspectratio=true,height=80mm]{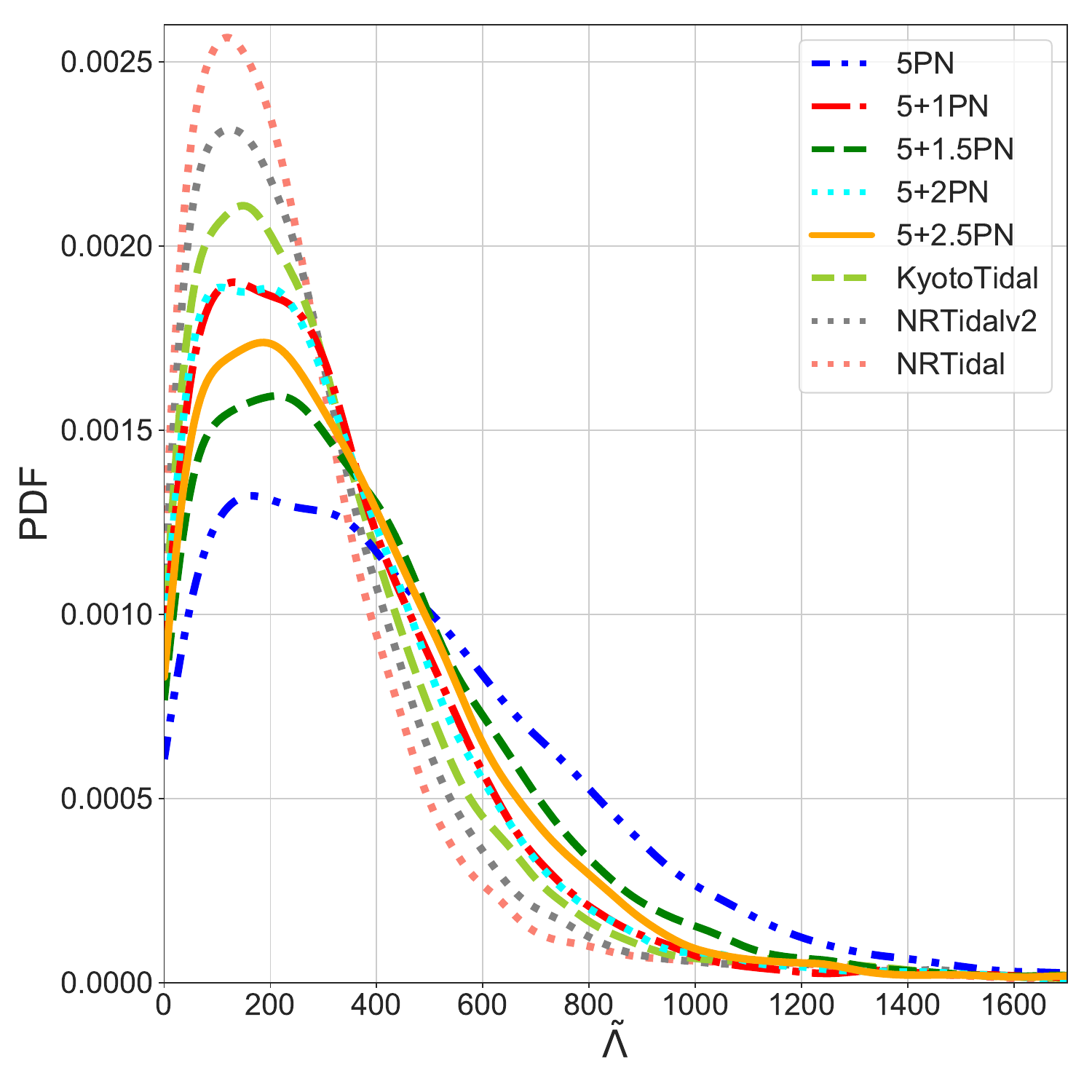}\\
 \end{center}
 \end{minipage}
\end{tabular}
  \caption{
Same as Fig.~\ref{fig:Lamt_GW170817_GW190425_PPs} but
estimated by using the point-particle model \texttt{TF2+} in common.
The tidal phasing is varied from 5PN through 5+2.5PN in the \texttt{PNTidal} model.
For comparison, the results obtained by using the NR calibrated models: the \texttt{KyotoTidal}, \texttt{NRTidalv2}, and \texttt{NRTidal} models, are also shown.
Here the BBH baseline is the \texttt{TF2+} model, $|\chi_{1z,2z}| \leq 0.05$ and the upper frequency cutoff $f_\mathrm{high}=1000~\mathrm{Hz}$ are set.
Narrow posterior uncertainties for the NR calibrated models compared to the \texttt{PNTidal} shows that NR information improves the tidal inference.
}
\label{fig:Lamt_GW170817_GW190425_differentPN_Kyoto_NRTidal}
\end{center}
\end{figure*}

\begin{figure*}[htbp]
  \begin{center}
\begin{tabular}{lcr}
 \begin{minipage}[h]{0.4\linewidth}
 \begin{center}
    \includegraphics[keepaspectratio=true,height=70mm]{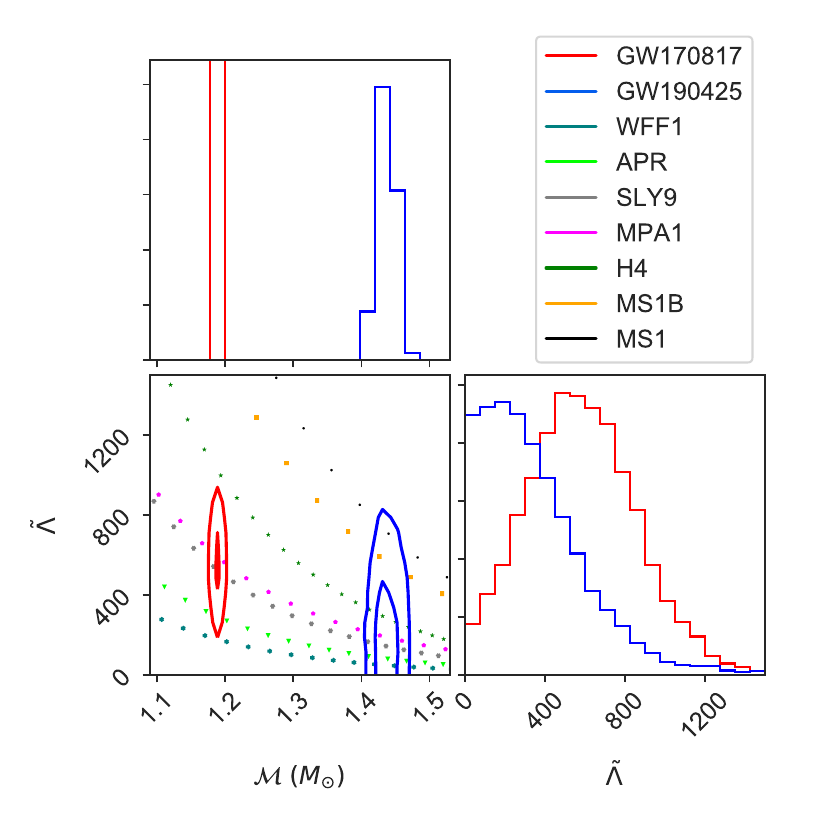}\\
 \end{center}
 \end{minipage}
 \begin{minipage}[h]{0.4\linewidth}
  \begin{center}
    \includegraphics[keepaspectratio=true,height=70mm]{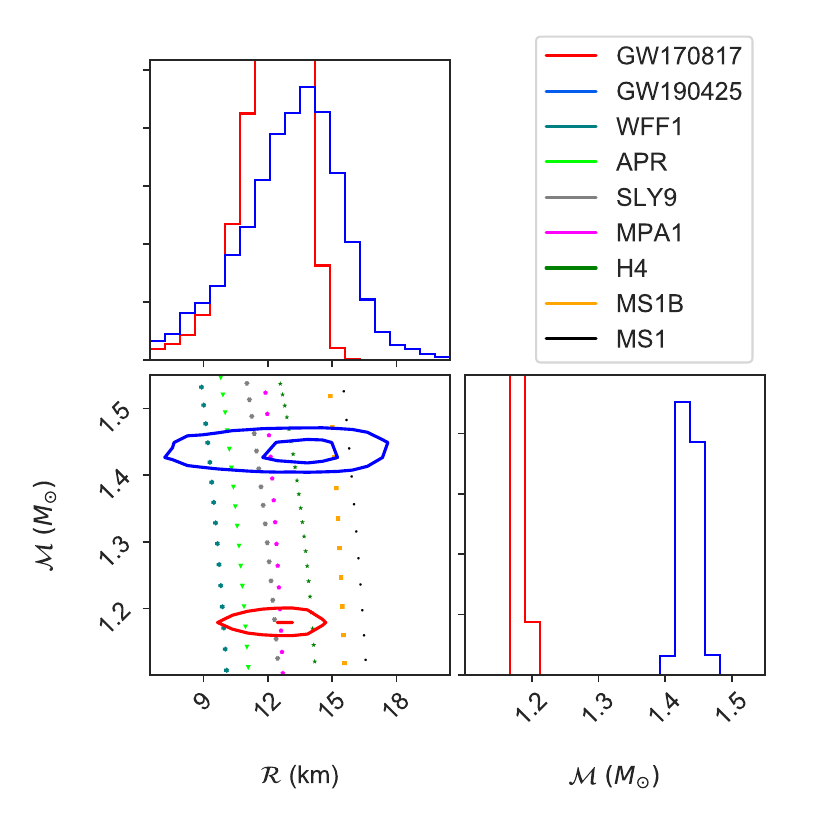}\\
     \end{center}
 \end{minipage}
 \begin{minipage}[h]{0.2\linewidth}
  \begin{center}
    \includegraphics[keepaspectratio=true,height=40mm]{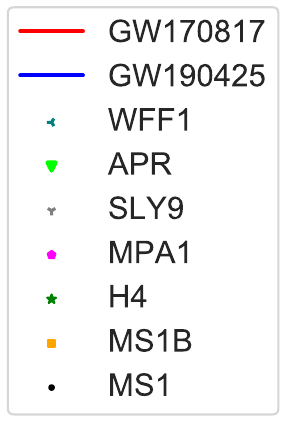}
     \end{center}
 \end{minipage}
\end{tabular}
  \caption{
Posterior PDFs of $\tilde{\Lambda}$-$\mathcal{M}$ (left) and $\mathcal{M}$-$\mathcal{R}$ (right) plane compared across posteriors and EOS model predictions:
50\% and 90\% contours show constraints obtained from GW170817 (red) and GW190425 (blue) by using the \texttt{TF2+\_PNTidal} model for $|\chi_{1z,2z}| \leq 0.05$ and the upper frequency cutoff $f_\mathrm{high}=1000~\mathrm{Hz}$.
The dots are selected EOS predictions (different colors and symbols), which correspond to binaries with equal masses.
Models that predict larger $\tilde{\Lambda}$: MS1 (black, point), MS1B (orange, square), and H4 (green, star) 
lie outside 90\% credible regions of $\tilde{\Lambda}$-$\mathcal{M}$ for GW170817.
}%
\label{fig:McLamt_RcMc_GW170817_GW190425}
\end{center}
\end{figure*}

\end{widetext}

\begin{table}[htbp]
\begin{center}
\caption{
The log Bayes factor and the associated standard deviation of the \texttt{TF2+\_PNTidal} model relative to the other point-particle baseline models, the \texttt{TF2g\_PNTidal} and \texttt{TF2\_PNTidal} models, 
$\log\mathrm{BF}_{\texttt{TF2+\_PNTidal}/\texttt{TF2g\_PNTidal},~\texttt{TF2\_PNTidal}}$, for GW170817 and GW190425. 
}
\vspace{5pt}
\begin{tabular}{lccc}
\hline \hline
Waveform & ~~GW170817 &  ~~GW190425 \\ \hline
\texttt{TF2g\_PNTidal} & $0.09_{-0.16}^{+0.16}$ & $0.12_{-0.15}^{+0.15}$ \\   
\texttt{TF2\_PNTidal} & $-0.20_{-0.16}^{+0.16}$ & $-0.18_{-0.15}^{+0.15}$ \\   
\hline \hline
\end{tabular}
\label{table:logBF_GW170817_GW190425_PPs}
\end{center}
\end{table}

\begin{table}[htbp]
\begin{center}
\caption{
The inferred marginalized posterior information of $\tilde{\Lambda}$ for GW170817 and GW190425 with different PN orders for \texttt{PNTidal}.
Here, the BBH baseline is the \texttt{TF2+} model. 
The medians with both 90\% symmetric and the HPD credible regions are reported.
}
\vspace{5pt}
\begin{tabular}{lccccc}
\hline \hline
 & ~~GW170817 &&  ~~GW190425 &\\ \hline
Waveform & ~~Symmetric & ~~HPD & ~~Symmetric & ~~HPD \\ \hline
5PN & $790_{-590}^{+664}$ & $790_{-632}^{+607}$ & $390_{-348}^{+743}$ & $390_{-389}^{+548}$ \\
5+1PN & $536_{-390}^{+436}$ & $536_{-419}^{+401}$ & $269_{-241}^{+545}$ & $269_{-269}^{+390}$ \\
5+1.5PN & $622_{-465}^{+529}$ & $622_{-499}^{+483}$ & $321_{-288}^{+639}$ & $321_{-321}^{+460}$ \\
5+2PN & $529_{-385}^{+447}$ & $529_{-424}^{+402}$ & $266_{-239}^{+567}$ & $266_{-266}^{+393}$ \\
5+2.5PN & $574_{-425}^{+485}$ & $574_{-467}^{+433}$ & $295_{-265}^{+578}$ & $295_{-295}^{+423}$ \\
\hline \hline
\end{tabular}
\label{table:Lamt_GW170817_GW190425_TF2plus_differentPN}
\end{center}
\end{table}

\begin{table}[htbp]
\begin{center}
\caption{
The log Bayes factor and the associated standard deviation of the terms at 5+2.5PN relative to the other PN orders, $\log \mathrm{BF}_\mathrm{5+2.5PN/different~PN}$, for GW170817 and GW190425. 
}
\vspace{5pt}
\begin{tabular}{lccc}
\hline \hline
PN order & ~~GW170817 &  ~~GW190425 \\ \hline
5PN & $-0.17_{-0.16}^{+0.16}$ & $-0.40_{-0.15}^{+0.15}$ \\   
5+1PN & $0.20_{-0.16}^{+0.16}$ & $0.14_{-0.15}^{+0.15}$ \\   
5+1.5PN & $-0.04_{-0.16}^{+0.16}$ & $-0.03_{-0.15}^{+0.15}$ \\   
5+2PN & $0.09_{-0.16}^{+0.16}$ & $0.02_{-0.15}^{+0.15}$ \\ 
\hline \hline
\end{tabular}
\label{table:logBF_GW170817_GW190425_TF2plus_differentPN}
\end{center}
\end{table}

\begin{table}[htbp]
\begin{center}
\caption{
The inferred marginalized posterior information of $\tilde{\Lambda}$ for GW170817 and GW190425 with various waveform models.
Here, the BBH baseline is the \texttt{TF2+} model 
and the $\tilde{\Lambda}$-form defined as Eq.~(\ref{eq:NUTTidal_phase_LamtForm}) is used
to take into account the prior volume reduction.
The medians with both 90\% symmetric and the HPD credible regions are reported.
}
\vspace{5pt}
\begin{tabular}{lccccc}
\hline \hline
 & ~~GW170817 &&  ~~GW190425 &\\ \hline
Waveform & ~~Symmetric & ~~HPD & ~~Symmetric & ~~HPD \\ \hline
\texttt{PNTidal} & $576_{-429}^{+480}$ & $576_{-468}^{+424}$ & $294_{-264}^{+599}$ & $294_{-294}^{+424}$ \\
\texttt{KyotoTidal}  & $490_{-364}^{+417}$ & $490_{-403}^{+366}$ & $247_{-219}^{+567}$ & $247_{-247}^{+386}$ \\
\texttt{NRTidalv2} & $446_{-322}^{+391}$ & $446_{-359}^{+349}$ & $223_{-199}^{+527}$ & $223_{-223}^{+351}$ \\
\texttt{NRTidal} & $413_{-306}^{+376}$ & $413_{-340}^{+330}$ & $202_{-180}^{+527}$ & $202_{-202}^{+328}$ \\
\hline \hline
\end{tabular}
\label{table:Lamt_GW170817_GW190425_TF2plus_Kyoto_NRTidal}
\end{center}
\end{table}

\begin{table}[htbp]
\begin{center}
\caption{
The log Bayes factor and the associated standard deviation of \texttt{PNTidal}
relative to the NR calibrated models: the \texttt{KyotoTidal}, \texttt{NRTidalv2}, and \texttt{NRTidal} models,
$\log\mathrm{BF}_{\texttt{PNTidal}/\mathrm{NR~calibrated~model}}$, 
for GW170817 and GW190425. 
Here, the $\tilde{\Lambda}$-form defined as Eq.~(\ref{eq:NUTTidal_phase_LamtForm}) is used for the \texttt{PNTidal} model to take into account the prior volume reduction.
In the last row, we show the values for the \texttt{TF2+} model as a BBH (nontidal) for comparison.
The values indicate no preference model on GW170817.
}
\vspace{5pt}
\begin{tabular}{lccc}
\hline \hline
Waveform & ~~GW170817 &  ~~GW190425 \\ \hline
\texttt{KyotoTidal} & $0.25_{-0.14}^{+0.14}$ & $0.22_{-0.14}^{+0.14}$ \\
\texttt{NRTidalv2} & $0.23_{-0.14}^{+0.14}$ & $0.32_{-0.14}^{+0.14}$ \\
\texttt{NRTidal} & $0.46_{-0.14}^{+0.14}$ & $0.37_{-0.14}^{+0.14}$ \\
BBH (nontidal) & $0.79_{-0.13}^{+0.13}$ & $-1.58_{-0.14}^{+0.14}$ \\
\hline \hline
\end{tabular}
\label{table:logBF_GW170817_GW190425_TF2plus_Kyoto_NRTidal}
\end{center}
\end{table}

\section{Conclusion}
\label{sec:summary}
Recently, PN tidal phase has been completed for 5+2PN order terms and corrected for 5+2.5PN order terms.
We present the results for the reanalyses of the BNS signals GW170817 and GW190425 
by using the PN tidal waveform model \texttt{PNTidal}:
estimates on the binary tidal deformability $\tilde{\Lambda}$
and constraints on EOS models for NSs.
To restrict to the inspiral regime, we set the upper frequency cutoff $f_\mathrm{high}=1000~\mathrm{Hz}$.

As a sanity check, we compare the results for the BNS events obtained by using the old and corrected versions of the \texttt{PNTidal} model employing the same point-particle model \texttt{TF2+}.
Since the correction is numerically very small,
a large difference in the inference of the BNS signals GW170817 and GW190425 
between the old and corrected PN tidal phase is not expected.
As expected, we confirm that our corrected result agrees well with the previous result obtained with the old PN tidal phase model.
However, the incorrect one should not be used.

First, we study a potential systematic difference in estimates of $\tilde{\Lambda}$ for GW170817 and GW190425 among different descriptions for the point-particle part.
We compare the results obtained with three PN models for the point-particle part, \texttt{TF2}, \texttt{TF2g}, and \texttt{TF2+}, employing the same tidal model \texttt{PNTidal}.
We find the absence of significant systematic difference in the estimates of $\tilde{\Lambda}$ among three PN point-particle models.
The Bayes factors indicate no preference among the different point-particle models for GW170817 and GW190425.

Then, by varying the PN orders from 5PN to 5+2.5PN for the tidal phase in the \texttt{PNTidal} model for our analyses, 
we investigate the effect of each PN order term on the estimate of $\tilde{\Lambda}$.
The estimates of $\tilde{\Lambda}$ slightly depend on the PN orders in the tidal phase.
We find that the estimates of $\tilde{\Lambda}$ does not monotonically change as the PN order increases 
and these results are well understood by comparing with the tidal phase shift for the different PN orders.
We also compare the results obtained by using the \texttt{PNTidal} and the NR calibrated tidal models: the \texttt{KyotoTidal}, \texttt{NRTidalv2}, and \texttt{NRTidal} models.
Here, they employ the same point-particle model \texttt{TF2+}.
Narrow posterior uncertainties for the NR calibrated models compared to the \texttt{PNTidal} shows that NR information improves inference of the tidal deformability.
We also find that the posterior PDFs of $\tilde{\Lambda}$ estimated by using the terms at 5+1PN and 5+2PN orders are
closer to the NR calibrated models than the terms at the half-PN orders: 5+1.5PN and 5+2.5PN orders.
These results are consistent with the tidal phase shift.
We compare the models with the Bayes factor.
The absolute magnitudes of the log Bayes factors between models are always less than 1,
which indicate no preference among the different PN orders nor the NR calibrated models over the \texttt{PNTidal} model 
by relying on the two BNS signals.

Finally, we present constraints on EOS models for NSs by combining information obtained from GW170817 and GW190425.
Our constraints with the \texttt{TF2+\_PNTidal} model for $f_\mathrm{high}=1000~\mathrm{Hz}$ show that GW170817 disfavor less compact EOS models,
which are consistent with the previous constraints obtained by the NR calibrated models for $f_\mathrm{high}=2048~\mathrm{Hz}$.
The 90\% allowed range of the chirp radius given by \texttt{TF2+\_PNTidal} is $9~\mathrm{km}\lesssim\mathcal{R}\lesssim15~\mathrm{km}$ for GW170817,
which is slightly weaker constraint than the \texttt{TF2+\_KyotoTidal} model.

For GW170817, with low SNR of $\sim30$, we find no significant systematic difference in the estimates of $\tilde{\Lambda}$ among PN point-particle models or PN and NR calibrated tidal models.
This means that the PN models work as a good approximation for the current detected BNS events, 
which is also confirmed using the Fisher information matrix formalism in Refs. \cite{Damour:2010zb, Chatziioannou:2017tdw}.
The KAGRA detector has recently joined the international network of GW detectors~\cite{LIGOScientific:2022myk, KAGRA:2022fgc}
and the Advanced LIGO and Advanced Virgo detectors are improving their sensitivities now.
They will detect BNS signals with high SNR and provide more information on the sources in coming observation runs~\cite{KAGRA:2013rdx}.
As the number of BNS coalescence events increases
and sensitivities of detectors are improved, 
the systematic differences among different point-particle models and tidal effects
will become significant 
and constraints on EOS models for NSs will significantly improved
as shown in Refs.~\cite{DelPozzo:2013ala, Agathos:2015uaa, Lackey:2014fwa, Wysocki:2020myz, Dudi:2018jzn, Samajdar:2018dcx, Messina:2019uby, Samajdar:2019ulq, Agathos:2019sah, Gamba:2020wgg, Landry:2020vaw, Chen:2020fzm, Chatziioannou:2021tdi, Kunert:2021hgm}.
Therefore, future data will reveal importance of extended waveform models from the PN models.


\section*{Acknowledgment}
We would like to thank Kyohei Kawaguchi, Hideyuki Tagoshi, and Wolfgang Kastaun for fruitful discussions and useful comments on the study.
This work is supported by JSPS KAKENHI Grants No. JP21K03548, No.~JP17H06361, No.~JP17H06358, and No.~JP17H06357.
We would also like to thank Computing Infrastructure ORION in Osaka City University and VELA in ICRR. 
This research has made use of data or software obtained from the Gravitational Wave Open Science Center (gw-openscience.org), a service of LIGO Laboratory, the LIGO Scientific Collaboration, the Virgo Collaboration, and KAGRA. LIGO Laboratory and Advanced LIGO are funded by the United States National Science Foundation (NSF) as well as the Science and Technology Facilities Council (STFC) of the United Kingdom, the Max-Planck-Society (MPS), and the State of Niedersachsen/Germany for support of the construction of Advanced LIGO and construction and operation of the GEO\,600 detector. Additional support for Advanced LIGO was provided by the Australian Research Council. Virgo is funded, through the European Gravitational Observatory (EGO), by the French Centre National de Recherche Scientifique (CNRS), the Italian Istituto Nazionale di Fisica Nucleare (INFN) and the Dutch Nikhef, with contributions by institutions from Belgium, Germany, Greece, Hungary, Ireland, Japan, Monaco, Poland, Portugal, Spain. The construction and operation of KAGRA are funded by Ministry of Education, Culture, Sports, Science and Technology (MEXT), and Japan Society for the Promotion of Science (JSPS), National Research Foundation (NRF) and Ministry of Science and ICT (MSIT) in Korea, Academia Sinica (AS) and the Ministry of Science and Technology (MoST) in Taiwan.

\appendix

\section{Comparison between the old and corrected PN tidal phases}
\label{sec:Old_vs_Corrected_PNTidal}
\begin{figure}[htbp]
  \begin{center}
 \begin{center}
    \includegraphics[keepaspectratio=true,height=70mm]{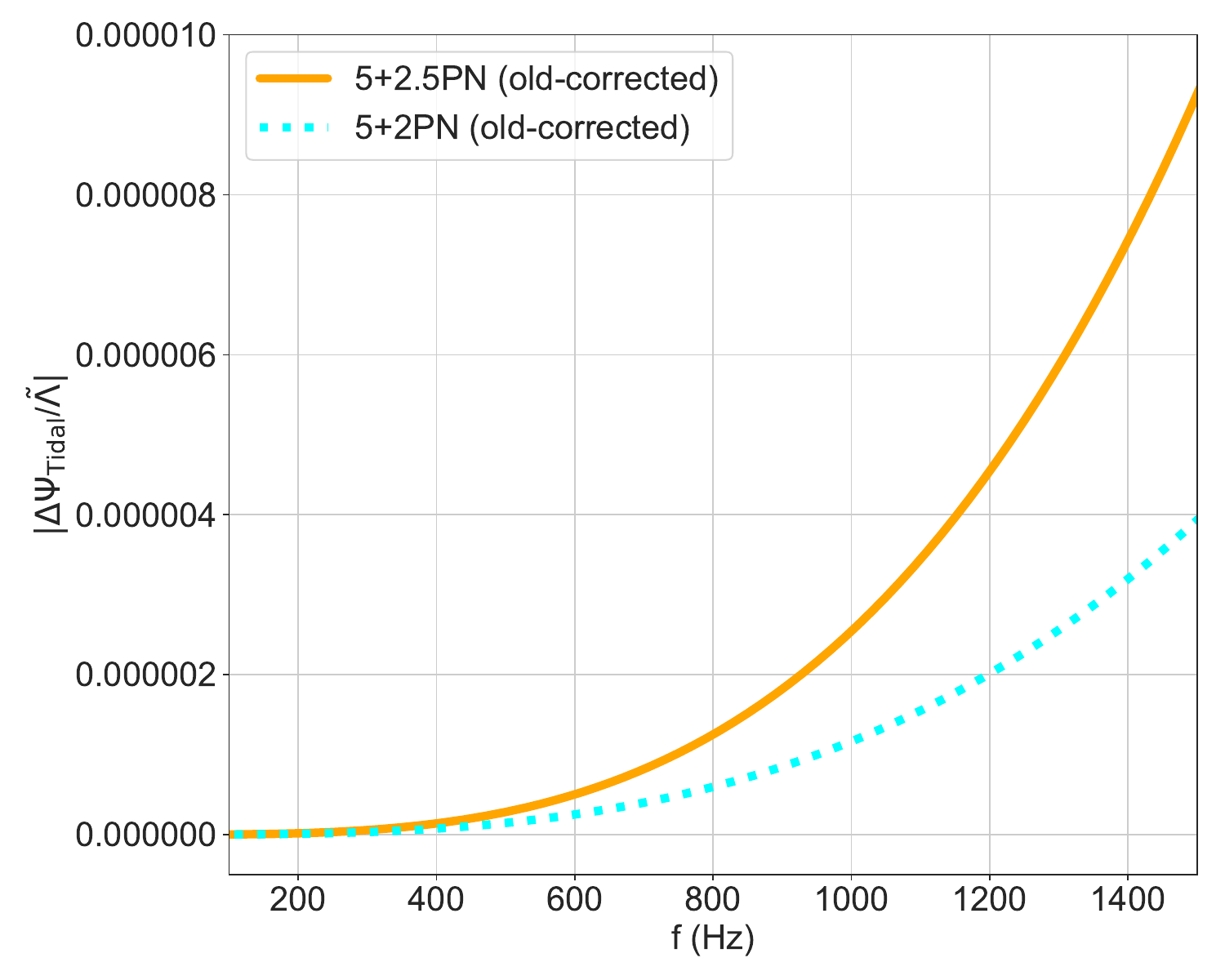}\\
 \end{center}
  \caption{
Absolute magnitude of the tidal phase difference between the old and corrected versions of \texttt{PNTidal} 
for 5+2PN (orange, solid) and 5+2.5PN (cyan, dotted) orders.
Here, they are divided by $\tilde{\Lambda}$ and calculated for the same unequal-mass case as used in Fig.~\ref{fig:DiffPhase_PP}.
The difference between the old and corrected versions are numerically very small.
}%
\label{fig:PhaseDifference_PNTidal}
\end{center}
\end{figure}

Here, we summarize phase difference between the old and corrected versions of the tidal phase in the \texttt{PNTidal} model.
Figure~\ref{fig:PhaseDifference_PNTidal} shows absolute magnitude of the tidal phase difference between the old and corrected versions of the \texttt{PNTidal} model for the 5+2PN and 5+2.5PN orders.
Here, they are divided by $\tilde{\Lambda}$ and calculated for the same unequal-mass case as used in Fig.~\ref{fig:DiffPhase_PP}.
This figure shows that the difference between old and corrected versions are numerically very small.

\section{Waveform systematics for the point-particle part in the source properties}
\label{sec:WFsys}
In this appendix, we present estimates of source parameters for completeness obtained by using three models for a point-particle part, the \texttt{TF2}, \texttt{TF2g}, and \texttt{TF2+} models, 
employing the same tidal model \texttt{PNTidal} for $|\chi_{1z,2z}| \leq 0.05$ and the upper frequency cutoff $f_\mathrm{high}=1000~\mathrm{Hz}$.
We demonstrate the results are robust to systematic uncertainty in the PN point-particle waveform models.

Figures~\ref{fig:MtotMcqEtaChieffLamt_GW170817} and \ref{fig:MtotMcqEtaChieffLamt_GW190425} 
show two-dimensional posterior PDFs of ($M^\mathrm{det}$, $\mathcal{M}^\mathrm{det}$, $q$, $\eta$, $\chi_\mathrm{eff}$, $\tilde{\Lambda}$) for GW170817 and GW190425, respectively.
The corresponding 90\% intervals for the \texttt{TF2+\_PNTidal} model are shown in Table~\ref{table:all_GW170817_GW190425_TF2plus_PNTidal}.
The estimates of source parameters presented
show the absence of significant systematic difference 
associated with a difference among PN waveform models for point-particle parts: the \texttt{TF2}, \texttt{TF2g}, and \texttt{TF2+} models.

\begin{widetext}

\begin{figure}[htbp]
  \begin{center}
 \begin{center}
    \includegraphics[keepaspectratio=true,height=160mm]{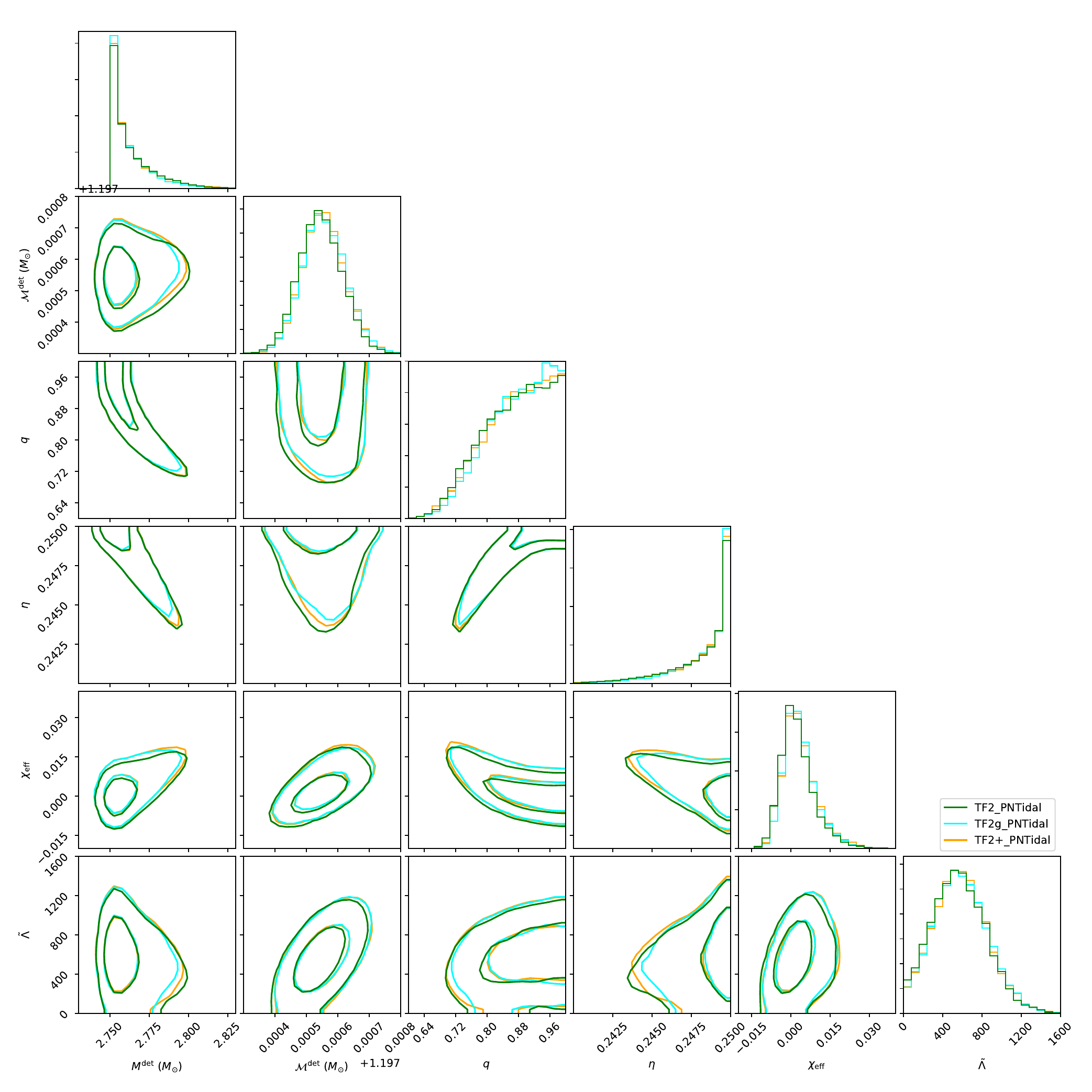}\\
 \end{center}
  \caption{
Comparison of two-dimensional posterior PDFs of ($M^\mathrm{det}$, $\mathcal{M}^\mathrm{det}$, $q$, $\eta$, $\chi_\mathrm{eff}$, $\tilde{\Lambda}$) for GW170817.
Contours show 50\% and 90\% credible regions for the \texttt{TF2\_PNTidal} (green), \texttt{TF2g\_PNTidal} (cyan), and \texttt{TF2+\_PNTidal} (orange) models for $|\chi_{1z,2z}| \leq 0.05$ and the upper frequency cutoff $f_\mathrm{high}=1000~\mathrm{Hz}$.
The systematic difference associated with a difference among waveform models for point-particle part 
are very small.
}%
\label{fig:MtotMcqEtaChieffLamt_GW170817}
\end{center}
\end{figure}

\begin{figure}[htbp]
  \begin{center}
 \begin{center}
    \includegraphics[keepaspectratio=true,height=160mm]{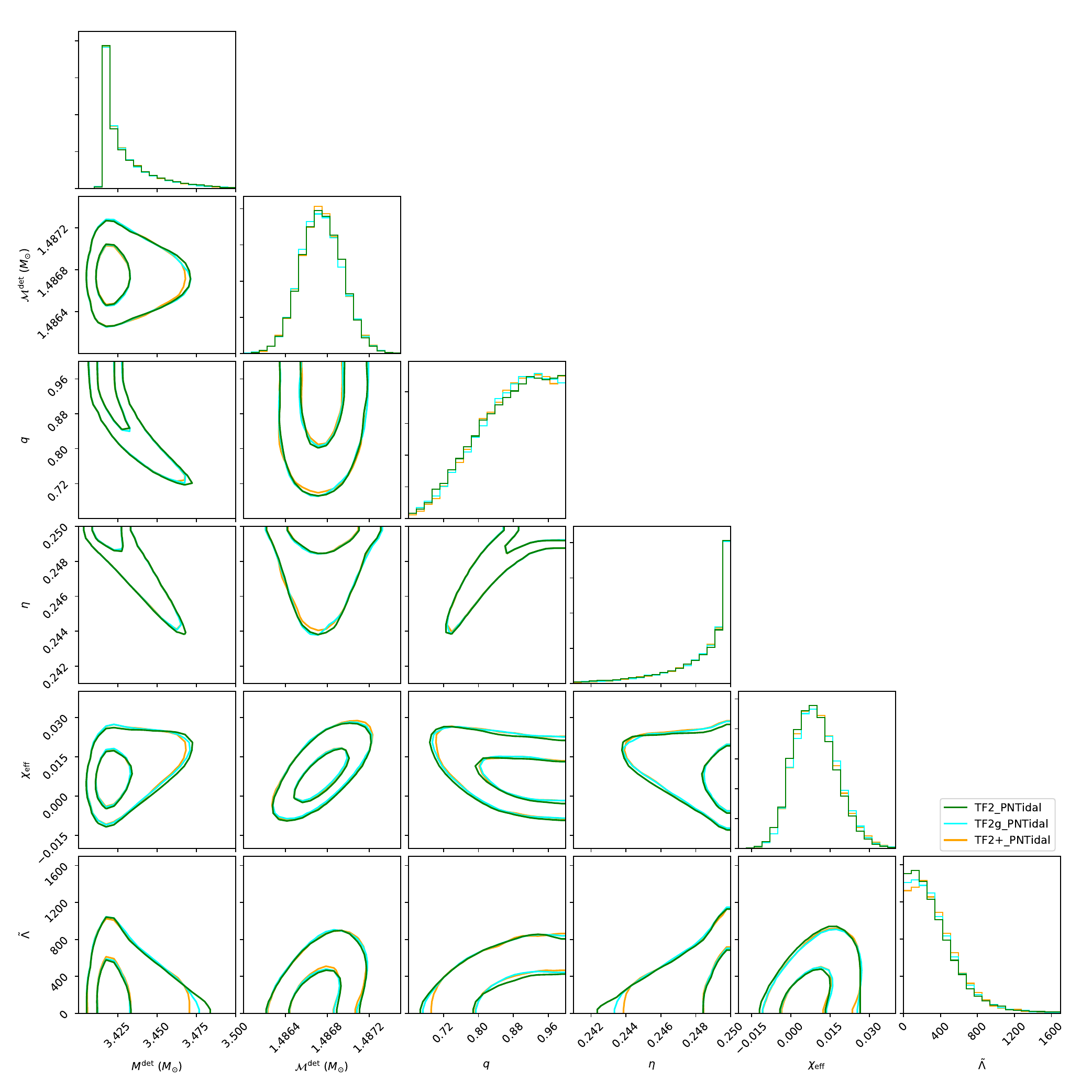}\\
 \end{center}
  \caption{
Same as Fig.~\ref{fig:MtotMcqEtaChieffLamt_GW170817} but for GW190425.
The systematic difference associated with a difference among waveform models for point-particle part 
are very small.
}%
\label{fig:MtotMcqEtaChieffLamt_GW190425}
\end{center}
\end{figure}
\end{widetext}

  

\end{document}